\documentclass[apj]{emulateapj}
\usepackage[colorlinks,linkcolor={blue},citecolor={blue},urlcolor={red}]{hyperref}
\usepackage{epsfig,graphicx,natbib,amsmath,amsfonts,amssymb}
\usepackage{color}
\usepackage{multirow}
\usepackage{booktabs}
\usepackage{amssymb}
\usepackage{threeparttablex} 
\usepackage{booktabs} 

\usepackage{longtable}
\usepackage{array} 

\newcommand{\sgm}{$\Sigma_{\rm 1kpc}$}

\newcommand{\msolar}{M$_\odot$}
\newcommand{\mstar}{$M_\ast$}
\newcommand{\lgmstar}{$\log_{10}$($M_\ast$/\msolar)}

\newcommand{\zgas}{12$+\log$(O/H)}

\newcommand{\nuvr}{NUV$-r$}

\newcommand{\sersic}{S\'{e}rsic}

\newcommand{\HI}{H{\sc{I}}}

\newcommand{\myemail}{\email{ecwang16@ustc.edu.cn(EW); xkong@ustc.edu.cn(XK)}}

\shorttitle{Connecting compact and extended star-forming galaxies}
\shortauthors{Wang et al.}

\graphicspath{{fig/}}

\begin{document}

\title{Connecting compact star-forming and extended star-forming galaxies at low-redshift: implications for galaxy compaction and quenching}
\author{
Enci Wang\altaffilmark{1,2,3},
Xu Kong\altaffilmark{1,2},
Zhizheng Pan\altaffilmark{4}
}\myemail

\altaffiltext{1}{CAS Key Laboratory for Research in Galaxies and Cosmology, Department of Astronomy, University of Science and Technology of China, Hefei 230026, China}
\altaffiltext{2}{School of Astronomy and Space Science, University of Science and Technology of China, Hefei 230026, China}
\altaffiltext{3}{Department of Physics, Institute for Astronomy, ETH Zurich, CH-8093 Zurich, Switzerland}
\altaffiltext{4}{Purple Mountain Observatory, Chinese Academy of Sciences, 2 West-Beijing Road, Nanjing 210008, China}

\begin{abstract}  
Previous findings show that the existence of dense cores or bulges is the prerequisite for quenching a galaxy,  leading to a  proposed two-step quenching scenario: compaction and quenching. In this scenario, galaxies first grow their cores to a stellar mass surface density threshold and subsequently quenching occurs, suggesting that galaxies evolve from extended star-forming galaxies (eSFGs), through compact star-forming galaxies (cSFGs), to quenched population. 
In this work, we aim at examining the possible evolutionary link between eSFGs and cSFGs by identifying the trends in star formation rate (SFR), gas-phase metallicity and \HI\ content, since one would naturally expect that galaxies evolve along the track of cold gas consumption and metal enhancement. 
We select a volume-limited sample of 15,933 galaxies with stellar mass above $10^{9.5}$\msolar\ and redshift of $0.02<z<0.05$ from the NASA-Sloan-Atlas catalog within the ALFALFA footprint.
The cSFGs on average exhibit similar or slightly higher SFRs of $\sim$0.06 dex and significantly higher gas-phase metallicity (up to 0.2 dex at low mass) with respect to the eSFGs, while the cSFGs dominate the galaxy population of the most intense star formation activities. 
More importantly, overall the median \HI\ content and gas depletion time of cSFGs are about half of eSFGs. 
Our result supports the compaction and quenching scenario that galaxies evolve and grow their cores along the track of  cold gas consumption and metal enhancement.
The environments of eSFGs and cSFGs are indistinguishable, suggesting that the compaction process is independent of any environmental effects at least for low-redshift universe. 

 
\end{abstract}


\section{Introduction}
\label{sec:introduction}

The bimodality of color or star formation rate for galaxies has been found in decades by large imaging and spectroscopic surveys \citep[e.g.][]{Strateva-01, Baldry-04, Bell-04, Blanton-05, Faber-07, Wetzel-Tinker-Conroy-12}. Galaxies thus are naturally separated into two populations: the star-forming (SF) galaxies and quenched galaxies (QGs). The typical SF galaxies are actively forming stars with prominent disk-like morphology, whereas the quenched galaxies usually have no no-going star formation with pronounced spheroid-like morphology \citep[e.g.][]{Strateva-01, Kauffmann-03, Baldry-04, Brinchmann-04, Li-06, Muzzin-13, Barro-17, Wang-18a}. Deep narrow-field surveys show that this bimodality persists up to redshift of 2.5 \citep[e.g.][]{Bundy-06, Brammer-09, Huang-13}, and the prevalence of quenched population has dramatically increased since redshift of 1 \citep[e.g.][]{Muzzin-13, Tomczak-14}, indicating that quenching 
is one of the major themes in galaxy evolution over the past 8 Gyr. However, how SF galaxies become to be quenched is still not well understood. 

Many theoretical processes have been proposed to explain the star formation quenching, which can mainly be categorized into two classes: the internal processes and external processes.  The feedback from starbursts and active galactic nuclei \citep[AGN;][]{McNamara-00, Nulsen-05, McNamara-Nulsen-07, Dunn-10, Fabian-12, Cicone-14}, belongs to the former class, act to heat the surrounding gas and/or strip the cold gas away from host galaxies.  Another example of internal process is so-called ``morphological quenching" \citep{Martig-09}, i.e. the presence of a dominant bulge stabilizes the gas disk against gravitational instabilities needed for star formation.  The external processes include a series of the environmental effects, such as major/minor mergers \citep[e.g.][]{Conselice-03, Cox-06, Smethurst-15}, and tidal/ram-pressure stripping \citep{Gunn-Gott-72,Moore-96, Abadi-Moore-Bower-99, Poggianti-17}, shock-excited heating \citep[e.g.][]{Rees-Ostriker-77, Birnboim-Dekel-03, Keres-05, Cattaneo-06} and strangulation \citep[e.g.][]{Weinmann-09, Peng-Maiolino-Cochrane-15, vandeVoort-17}, which act to either rapid consume the cold gas and/or expel the cold gas, or prevent the cold gas accretion.   
However, some of these internal processes may be closely related to environmental effects, which makes the star formation cessation picture to be rather complicated. For instance, the AGN activities can be triggered by galaxy-galaxy interactions or mergers \citep{Urrutia-Lacy-Becker-08, Ellison-11, Kocevski-12, Satyapal-14}. 
Since all the quenching mechanisms are working on the cold gas of galaxies, understanding the cold gas content in galaxies is essential to uncover the stage of galaxy evolution along the track of star formation quenching. 

Observationally, the link between quenching and structural properties has caught attention. 
Many works proposed that massive bulge is the key factor for quenching star formation in local galaxies \citep[e.g.][]{Bell-12, Cheung-12, Wake-vanDokkum-Franx-12, Fang-13, Bluck-14, Bluck-16}. 
For instance, \cite{Fang-13} found that the existence of a dense bulge is necessary but not sufficient to quench a galaxy by analyzing the stellar surface density profiles for SF and quenched central galaxies.   
More recently, by applying an artificial neural network approach for pattern recognition of quiescent systems on  $\sim$400,000 central galaxies taken from SDSS, \cite{Teimoorinia-Bluck-Ellison-16} argued that the central velocity dispersion, bulge mass and the bulge-to-total stellar mass ratio are excellent quenching predictors, indicating that properties related to the central mass of the galaxy are most closely linked to the star formation cessation. 
The link between quenching and structural properties has also been established for high-redshift galaxies \citep[e.g.][]{Tacchella-15, Barro-17}. The fundamental structural differences for SF and quiescent galaxies are presented in the mass-size relation of different redshifts \citep[e.g.][]{Toft-07, Williams-10, Newman-12, vanderWel-14, Shibuya-Ouchi-Harikane-15}, where quiescent galaxies always exhibit a much higher stellar surface density than SF galaxies. These suggest that SF galaxies must grow dense cores before quenching.  

Recently, a two-step star formation quenching scenario has been proposed: compaction and quenching \citep[e.g.][]{Fang-13, Dekel-Burkert-14, Tacchella-15, Tacchella-16a, Tacchella-16b, Barro-17}. 
The term compaction, as used in the present work, can be caused by the shrinkage of galaxies with radial migration of stars and/or the growth of the cores or bulges without a significant change in overall radius (see Section \ref{subsec:compaction} for details). 
Under this scenario, in high redshift universe, the extended SF galaxies firstly contract via a dissipative process to loss energy and angular momentum, such as major merger, accretion of counter-rotating streams or recycled gas, usually associated with violent disc instability \citep{Dekel-Burkert-14, Zolotov-15}, subsequently become compact SF galaxies, then consume and/or lose their cold gas and turn to quiescent galaxies as a whole.  In low redshift universe, the compaction of extended SF galaxies may be via minor merger, interaction with neighboring galaxies and/or bar-driven secular evolution to form the compact SF galaxies \citep[e.g.][]{Moore-96, Moore-Lake-Katz-98, DiMatteo-07, Bournaud-Jog-Combes-07, Wang-12, Barro-17, Lin-17}. 
Observationally, there are indicative identifications of the progenitors of the compact quiescent galaxies in the form of compact SF galaxies, whose masses, kinematics and morphologies resemble those of the compact quiescent galaxies \citep{Barro-13a, Barro-14b, Bruce-14, Nelson-14, Williams-14}, but appear to be different from other SF galaxies that have irregular or clumpy features. However, whether the extended SF galaxies are necessarily to be quenched through compaction, how the compaction process occurs and what happens in this process are still poorly understood. 

In this paper, we aim at testing the two-step quenching scenario, by comparing the extended SF galaxies and compact SF galaxies to find if there are any clues for the supposed evolution from eSFGs and cSFGs under the two-step quenching scenario. Following the work of \cite{Fang-13}, we use the stellar surface mass density within a radius of 1 kpc, \sgm, to quantify the growth of bulge as galaxies evolve. They have presented an existence of \sgm\ threshold, above which galaxies begin to shut down their star formation. This result persists up to at least redshift of 3 \citep{Barro-17, Whitaker-17, Mosleh-17}. While there is also a significant fraction of SF galaxies with \sgm\ greater than the threshold, studying their properties and connecting them with extended SF galaxies would probably shed light on the process of compaction and quenching. 
Thus, in this work, we select the sample from NASA Sloan Atlas and separate them into extended SF galaxies and compact SF galaxies\footnote{This definition of compact SF galaxies differs from previous works, where compact is an absolute term to identify the smallest galaxies \citep[e.g.][]{vanDokkum-15}.} according to \sgm-\mstar\ diagram. Although the sample galaxies we selected are limited to the low-redshift universe, this enables us to present detailed investigation on the chemical abundance, \HI\ content, and the environment for eSFGs and cSFGs, and further to examine whether there is a trend for cold gas consumption and metal enhancement from eSFGs to cSFGs.

This paper is structured as follows. In Section \ref{sec:data}, we present the details on sample selection and  definition of extended SF and compact SF galaxies, as well as the description of relevant physical parameters. In Section \ref{sec:results}, we investigate the star formation rates, gas-phase metallicities, \HI\ detection rate and environments as a function of stellar mass for extended SF and compact SF galaxies. In Section \ref{sec:discussion}, we discuss the gas depletion time along the star formation main sequence, the prevalence of AGN for the two populations,  and the implications for our result. We summarize the result in Section \ref{sec:summary}. Throughout this paper, we assume a flat cold dark matter cosmology model with $\Omega_m=0.3$, $\Omega_\Lambda=0.7$ and $h=0.7$ when computing distance-dependent parameters.

\section{Data}
\label{sec:data}

\subsection{Sample selection}

\begin{figure*}
  \begin{center}
   \epsfig{figure=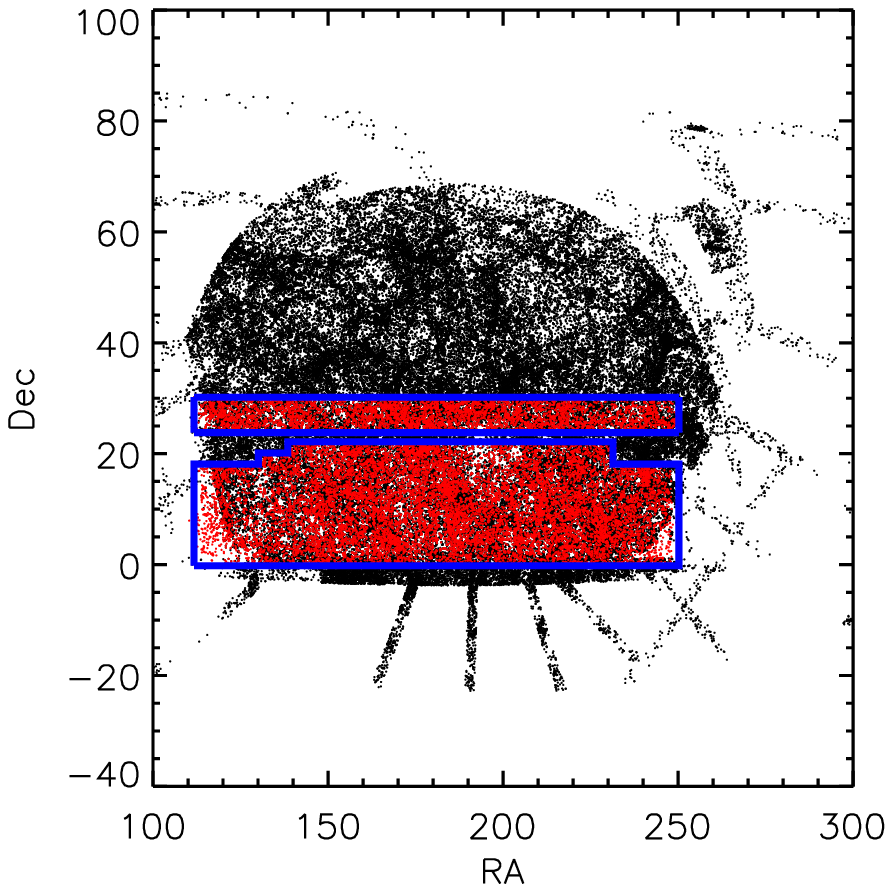,clip=true,width=0.38\textwidth}
   \epsfig{figure=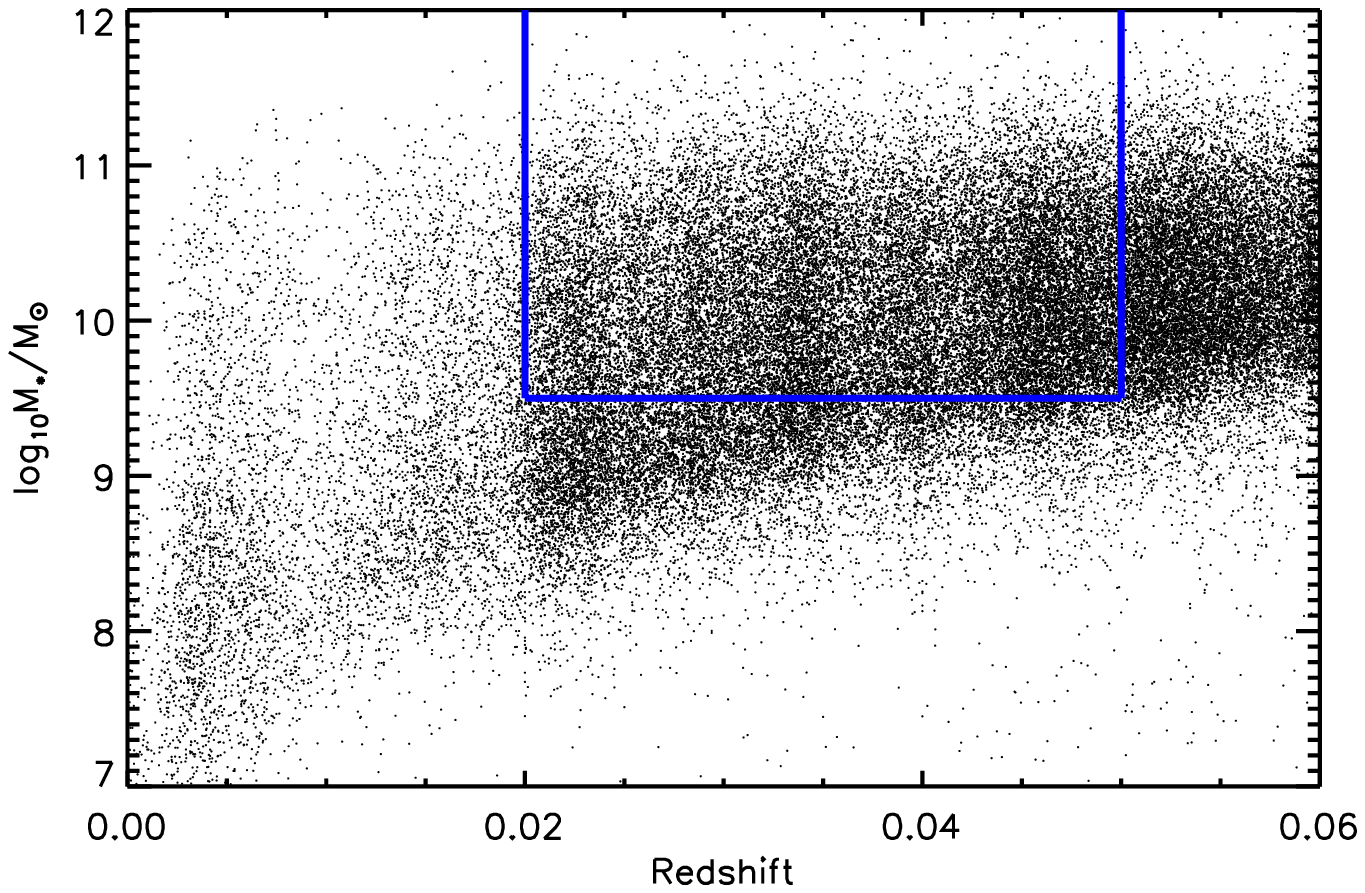,clip=true,width=0.57\textwidth}
  \end{center}
\caption{Left panel: the ALFALFA galaxy sample overlapped on the NSA footprint, indicated by each individual galaxies with redshift less than 0.06. The black small dots represent NSA galaxies sample, and the red small dots represent the ALFALFA sample. The blue lines indicate the boundaries of the ALFALFA footprint. Right panel: the stellar mass versus redshift relation for NSA sample within the ALFALFA footprint. We select a volume-limited galaxy sample with stellar mass greater than $10^{9.5}$\msolar\ and redshift of $0.02<z<0.05$, denoted by blue lines. }
 \label{fig:sample_selection}
\end{figure*}

Our galaxy sample is selected from the NASA Sloan Atlas (NSA)\footnote{http://www.nsatlas.org/},  which is a galaxy catalog constructed by \cite{Blanton-11} listing physical parameters for more than 640,000 local galaxies based on data from GALEX, SDSS and 2MASS. We select the galaxy sample to locate in the
ALFALFA\footnote{http://egg.astro.cornell.edu/index.php/} \citep[Arecibo Legacy Fast ALFA Survey; ][]{Giovanelli-05, Haynes-11} footprint. Thus, the \HI\ mass of galaxies can be obtained. The left panel of Figure \ref{fig:sample_selection} shows the ALFALFA galaxy sample (red small dots) overlapped on the NSA footprint, indicated by all the individual galaxies (black small dots). Here we only use the galaxies of the northern galactic cap from the 70\% ALFALFA catalog, since the southern galactic cap of ALFALFA has little overlap with the NSA footprint. The blue solid lines show the boundaries of ALFALFA footprint, and 258,938 NSA galaxies are within the boundaries. 

The right panel of Figure \ref{fig:sample_selection} presents the stellar mass and redshift relation for NSA sample within the ALFALFA footprint. According to this relation, we select a volume-limited sample with stellar mass greater than $10^{9.5}$\msolar, and redshift of $0.02<z<0.05$ to avoid the Malmquist bias, denoted by blue lines. The upper limit of redshift is selected by considering the redshift range of ALFALFA ($z<0.06$), and the lower limit of the stellar mass is selected to include more galaxies. In addition, we exclude the highly inclined galaxies with the minor-to-major axis ratio less than 0.5 to reduce the influence of inclination effect on the measurements of \sgm\ \citep{Fang-13}.  
According to the above criteria, in total 17,609 galaxies are selected. Since the ALFALFA is a blind \HI\ survey, we note that only a small fraction (27.5\%) of galaxies in the selected sample are detected in 21 cm emission by ALFALFA. We combine the selected sample with MPA-JHU catalog\footnote{http://wwwmpa.mpa-garching.mpg.de/SDSS/DR7} \citep{Brinchmann-04} to obtain the physical quantities of individual galaxies with excluding 9.5\% unmatched sources.
These unmatched sources are the galaxies in NSA catalog but not in MPA-JHU catalog, because the NSA catalog includes galaxies from a series of redshift surveys, such as SDSS DR8, NASA Extragalactic Database and Six-degree Field Galaxy Redshift Survey. 
Thus, our final sample includes 15,933 galaxies. 

\subsection{Physical properties of galaxies}
\label{subsec:parameters}

The MPA-JHU catalog provides the measurements of the main physical parameters used in this work, including star formation rate (SFR), gas-phase metallicity and the relevant emission line fluxes. The SFRs are measured by an updated version of the method of \cite{Brinchmann-04} using the Kroupa initial mass function \citep{Kroupa-Weidner-03}. The measurements of gas-phase metallicity are estimated using the \cite{Charlot-Longhetti-01} model, taken from  \cite{Tremonti-04}. 
The stellar masses are drawn from the NSA catalog. The surface mass density within a circular aperture of radius 1 kpc, denoted as \sgm, is calculated by directly integrating the light profiles from the innermost point out to 1 kpc, with adopting the relation between $M_*/L_i$ (stellar mass divided by $i$-band luminosity) and the rest-frame $g-i$ color from \cite{Fang-13}: $\log M_*/L_i=1.15+0.79\times(g-i)$.
In practice, we generate the cumulative flux profile at a series of radius and obtain the total flux within 1 kpc by the cubic spline interpolation for the two bands, based on the azimuthally-averaged radial surface brightness profile (also known as {\tt ProfMean} in the output of SDSS pipeline). 
The $i$-band luminosity and $g-i$ color are corrected to the rest-frame \citep{Blanton-Roweis-07}  with considering the galactic extinction \citep{Schlegel-Finkbeiner-Davis-98}. Our sample is selected to have redshift less than 0.05, where the half-width at half-maximum of the SDSS point-spread function (0.7 arcsec) is comparable to the aperture of 0.68 kpc. This indicates that the seeing should not heavily affect the reliability of measurements of \sgm\ \citep[see more details in][]{Fang-13}.  
Although the main result in this work is presented based on the \sgm, we have also adopted the stellar mass surface density within the effective radius to reproduce the main result and find that our result still holds (see Appendix A). This indicates that our result is not sensitive to the definition of the compactness of galaxies.  

Since only the most \HI-rich galaxies have been detected in 21 cm emission, we adopt an \HI\ mass estimator for the galaxies unmatched with ALFALFA, using the formula of \cite{Catinella-10}: $\log_{10}M_{\rm HI}/M_*=-0.332\times \mu_{*,z}-0.240\times({\rm NUV}-r)+2.856$, where $M_{\rm HI}$ is the \HI\ mass of galaxies, $\mu_{*,z}$ is the stellar mass surface density computed using the half-light radius of SDSS $z$-band images, and \nuvr\ is the color between near-UV and SDSS $r$-band. This relation is calibrated using the GASS \citep[The GALEX Arecibo SDSS Survey;][]{Catinella-10} sample with a scatter of $\sim$0.3 dex, and suggests that galaxies with higher $\mu_{*,z}$ and redder \nuvr\ color appear to be more \HI-deficient \citep{Brown-15}. 
\cite{Li-12} have found that this relation generally works well but underestimates the \HI\ mass for the most \HI-rich galaxies with  $\log_{10}M_{\rm HI}/M_*>$0.0. However, this should not be a concern, since the most \HI-rich galaxies have the observed \HI\ mass from ALFALFA and our purpose is to make comparison of compact SF and extended SF galaxies.   

We quantify the galaxy environment by using two independent-derived parameters: the host halo mass and the local overdensity of galaxies. The host halo masses are derived from the SDSS DR7 \citep{Abazajian-09} group catalog of \cite{Yang-07}, which is based on a halo-based group finding algorithm developed in \cite{Yang-05}.  The local overdensity ($\delta$) is defined as $\delta=\rho/\bar{\rho}$, where $\rho$ is the local matter density and $\bar{\rho}$ is the mean matter density. The local overdensity used here is drawn from the three-dimensional reconstructed local mass density of SDSS DR7 \citep{Abazajian-09} with a smoothed scale of 3 Mpc using the approach of non-linear, non-Gaussian full Bayesian large-scale structure analysis \citep{Jasche-10}. 

\subsection{Definition of cSFGs and eSFGs}

\begin{figure*}
  \begin{center}
     \epsfig{figure=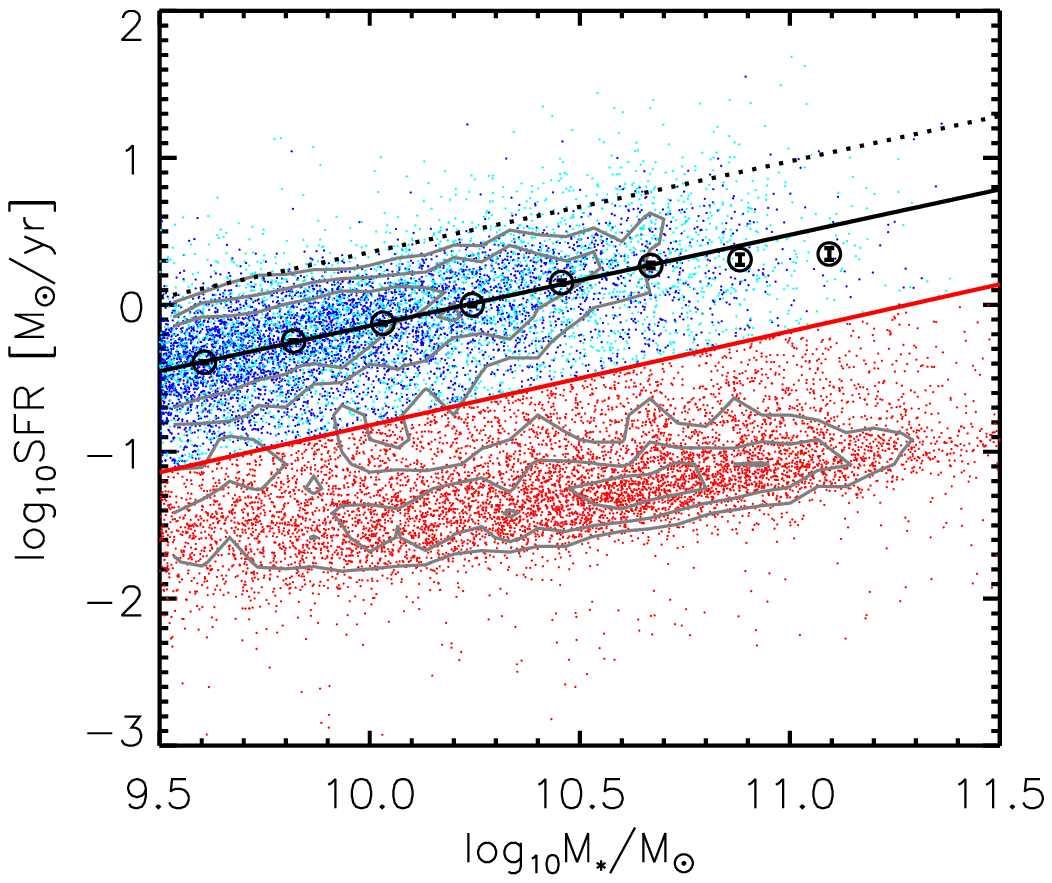,clip=true,width=0.45\textwidth}
   \epsfig{figure=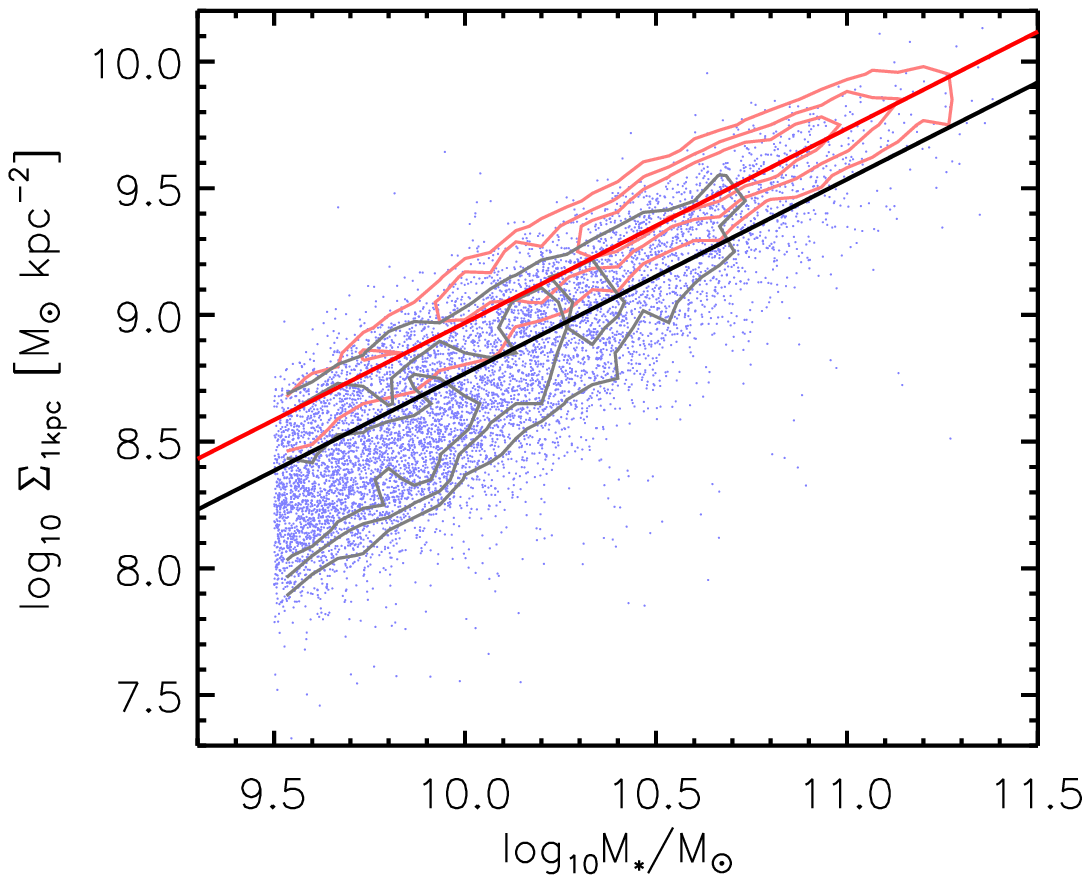,clip=true,width=0.45\textwidth}
  \end{center}
\caption{ Left panel: SFR versus stellar mass for cSFGs (cyan dots), eSFGs (blue dots) and quenched galaxies (red dots).  The red solid line is taken from \cite{Woo-13} to separate SF galaxies and quenched galaxies. The black circles show the median SFR of SF galaxies at given stellar mass bins, and the errors are estimated by the bootstrap method. 
The black solid line is the linear fitting of the black data points (except the two highest stellar mass bins), which is usually referred as normal SFMS. It can be written as: $\log_{10}{\rm SFR}=0.62\log_{10}M_{*}-6.32$.  
The black dotted line is 0.5 dex above the fitted SFMS. The gray contours indicate the distribution of galaxies on the SFR-$M_*$ diagram. 
Right panel: \sgm\ versus stellar mass for the sample galaxies.  The blue small dots represent the SF galaxies, overlapped with the gray contours.  The red contours represent the distribution of quenched galaxies on the \sgm-$M_*$ diagram.  The contour levels are corresponding to the 25\%, 50\% and 75\% of data points from inner outward for both gray and red contours.  
The red solid line is the best fit relation for quenched galaxies,  which is in the form as: $\log_{10}\Sigma_{\rm 1kpc}=0.77\log_{10}M_*+1.30$. The black solid line is parallel to but 0.2 dex lower than the red solid line, which is adopted to separate the cSFGs and eSFGs. }
 \label{fig:sample_definition}
\end{figure*}

Galaxies are bimodally distributed in the SFR-\mstar\ relation \citep[e.g.][]{Brinchmann-04, Peng-10, Woo-13, Bluck-16},  which is widely used to separate SF and quenched populations.  The left panel of Figure \ref{fig:sample_definition} shows the SFR-\mstar\ relation. As expected, the bimodal distribution of sample galaxies is clearly seen. We adopt the division line from \cite{Woo-13}: $\log_{10}{\rm SFR}=0.64\log_{10}M_{*}-7.22$, indicated by the red solid line, to separate SF galaxies and quenched galaxies. Thus, 8442 galaxies are classified SF galaxies, and 7491 galaxies are quenched ones.  According to this definition, some so-called ``green valley'' galaxies are classified into SF type.  We have checked our main result with shifting the demarcation line to avoid or largely reduce the mixing effect of green valley galaxies from SF ones, and find that the main result still holds. 
The SF galaxies lie in a tight sequence on the SFR-\mstar\ plane, which is usually known as the star formation main sequence (SFMS). For SF galaxies, we present the median SFR in a series of stellar mass bins, indicated by black circles. We perform a linear fit to the data points\footnote{In the fittings, we exclude two data points of highest stellar mass bins, since the SFMS becomes flatter at the high stellar mass end \citep[e.g.][]{Erfanianfar-16, Pan-Zheng-Kong-17}.}, shown in the black solid line. This black solid line thus represents the normal SFMS of the sample galaxies. 

The right panel of Figure \ref{fig:sample_definition} shows the \sgm-\mstar\ relation for SF galaxies (blue dots with gray contours) and quenched galaxies (red contours). As shown, a linear correlations are clearly seen for both SF  and quenched galaxies. The quenched galaxies reside in a narrow sequence on the \sgm-\mstar\ diagram, while SF galaxies appear to be more scatteredly distributed with respect to the quenched population. 
Furthermore, the quenched galaxies appear to have higher \sgm\ than SF galaxies across the whole stellar mass range. This is in good agreement with the previous finding that the existence of a mature bulge or a dense stellar core is prerequisite to quench a galaxy  \citep{Cheung-12, Fang-13, Bluck-14, Barro-17}.  
However, there is a significant fraction of SF galaxies that reside in the same region as quenched galaxies on the \sgm-\mstar\ diagram. These galaxies are likely in the transitional phase from normal SF galaxies to quenched galaxies according to the compaction and quenching scenario. Studying these galaxies would probably give instructions on galaxy compaction and star formation quenching. 
Thus, we separate SF galaxies into two subsamples: the compact SF galaxies (cSFGs) and extended SF galaxies (eSFGs). We emphasize that the definition of compact galaxies is not based on galaxy radius, but on the central surface stellar mass density throughout this paper.
The demarcation line is indicated by the black solid line, which is parallel to but 0.2 dex lower than the best-fit relation of \sgm-\mstar\ for quenched galaxies (the red solid line).  This demarcation is selected to be in line with the bottom envelop of the red contour enclosed 75\% of quenched population.  Although this demarcation is manually determined, it excellently distinguishes the \sersic\ index of cSFGs and eSFGs (see details in Appendix B), suggesting that the classification is reasonable.
Overall, the sample galaxies consists three subsamples: 3452 cSFGs, 4990 eSFGs and  7491 quenched galaxies. 

\section{Observational Results}
\label{sec:results}

The cSFGs and eSFGs are defined to have different inner stellar surface density. According to the compaction and quenching scenario, the eSFGs are expected to assemble their inner stellar mass before becoming quiescent galaxies. However, the comparison of physical properties for eSFGs and cSFGs are not well investigated up to now, such as their locations on the SFMS and the mass-metallicity relation, their cold gas content and environments. In this section, we will study the properties of these two populations, which would shed light on the galaxy compaction (if any) and further examine the compaction and quenching scenario. 


\subsection{Two approaches of galaxy compaction}
\label{subsec:compaction}

\begin{figure*}
  \begin{center}
    \epsfig{figure=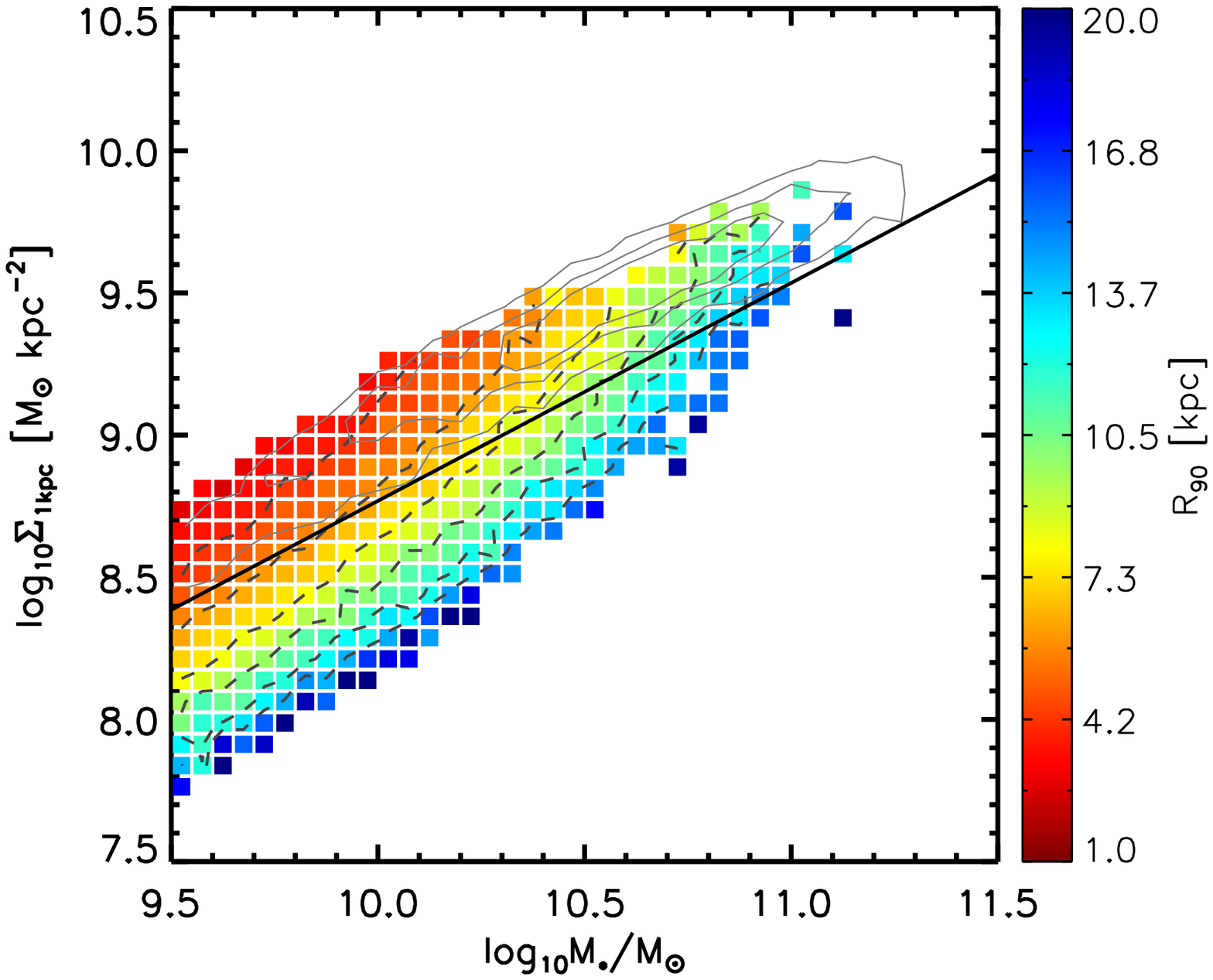,clip=true,width=0.405\textwidth}
   \epsfig{figure=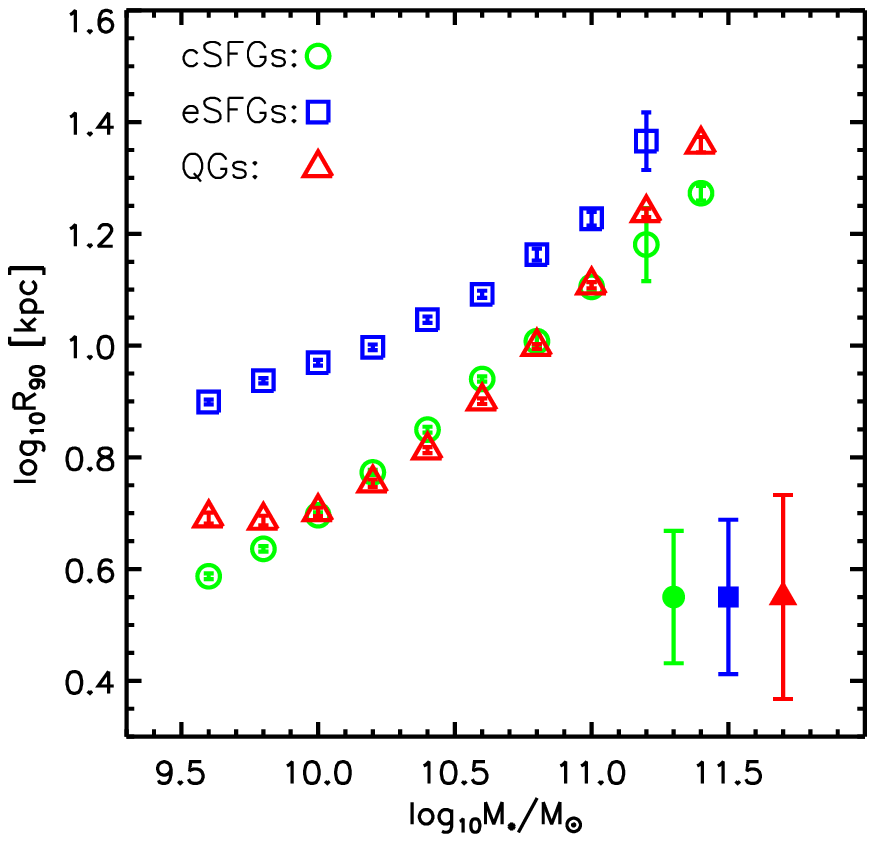,clip=true,width=0.378\textwidth}
  \end{center}
\caption{Left panel: the \sgm-\mstar\ diagram for SF galaxies color-coded with $R_{\rm r, 90}$. The gray solid contours indicate the distribution of quenched galaxies on \sgm-\mstar\ diagram, and the black solid line is the demarcation for cSFGs and eSFGs, taken from the right panel of Figure \ref{fig:sample_definition}.  The black dashed curves indicate the contours of constant $R_{\rm r, 90}$ on \sgm-\mstar\ diagram. 
In generating the plot, pixels with less than 3 galaxies are not shown, and we set the same limit when generating similar plots throughout the paper.  
Right panel: the median $R_{\rm r, 90}$ as a function of stellar mass for eSFGs (blue squares), cSFGs (green circles) and quenched galaxies (red triangles). The errors are estimated from the bootstrap method with 1000 times resamplings.  We also present the representative scatter of $R_{\rm r, 90}$-$M_*$ relation for eSFGs, cSFGs and QGs, denoted as three filled data points with large error bars in the bottom-right corner. The representative scatter is the median value of the scatters at different stellar masses.
}
\label{fig:size}
\end{figure*}

The SF galaxies and quenched galaxies are found to have different size both in low and high redshift \citep[e.g.][]{Toft-07, Williams-10, Newman-12, vanderWel-14, Shibuya-Ouchi-Harikane-15}, suggesting the co-evolution of the size and the star formation status of galaxies.    
The left panel of Figure \ref{fig:size} presents the \sgm-\mstar\ diagram of SF galaxies with the color-coding of the Petrosian radius ($R_{\rm r, 90}$), defined as the radius containing 90\% of the Petrosian luminosity based on SDSS $r$-band image. Here we use $R_{\rm r, 90}$ rather than half-light radius, because the $R_{\rm r, 90}$ is better representative of the global size of galaxies.  The black dashed curves indicate the contours of constant $R_{\rm r, 90}$ on the diagram.  
As expect, at fixed stellar mass, the $R_{\rm r, 90}$ is dramatically decreasing when galaxies are getting more compact, indicating that the cSFGs are much smaller in size than eSFGs at give stellar mass.  This result can be seen more clearly in the right panel of Figure \ref{fig:size}, where the $R_{\rm r, 90}$ as a function of stellar mass for eSFGs (blue squares), cSFGs (green circles) and quenched galaxies (red triangles) are shown. 
The representative scatter of the $R_{\rm r, 90}$-\mstar\ relation is denoted in the bottom right corner for each population.  As shown, eSFGs have systematically higher $R_{\rm r, 90}$ than cSFGs (for up to 0.3 dex), while the cSFGs show similar global size with quenched galaxies almost in the whole stellar mass range. 

Assuming that eSFGs need to evolve to cSFGs before quenching, eSFGs are becoming more and more compact mainly via two approaches: the in-situ star formation in their central regions, and/or the shrinkage of their sizes with the radial migration of stars.  These two different approaches result in different evolution tracks of eSFGs on the \sgm-\mstar\ diagram. 
For a given eSFG in the bottom left of \sgm-\mstar\ diagram, we naturally assume that it evolves both with increasing stellar mass and increasing \sgm\ as time goes on.  If the eSFG assembles its stellar mass only via in-situ star formation on its stellar disk, the global size should not be changed, and then it would evolve along the contours of constant $R_{\rm r, 90}$ (see the dashed curves in the left panel of Figure \ref{fig:size}). If the eSFG is getting compact mainly via the size shrinkage (possible along with in-situ star formation or mergers), it would evolve along a track that is steeper than the contours of constant $R_{\rm r, 90}$. In case the eSFG evolves along a track flatter than the contours of constant $R_{\rm r, 90}$, this means that its stellar mass and size are growing without prominent growing of \sgm, which is in good agreement with the inside-out growth scenario \citep[e.g.][]{Perez-13, Goddard-17, Wang-18b}.  
Thus, although the cSFGs are defined by using \sgm, the compaction process discussed in this work includes the above two approaches of compaction.  
 
\subsection{The SFR and gas-phase metallicity}
\label{subsec:sfr}

\begin{figure*}
 \begin{center}
   \epsfig{figure=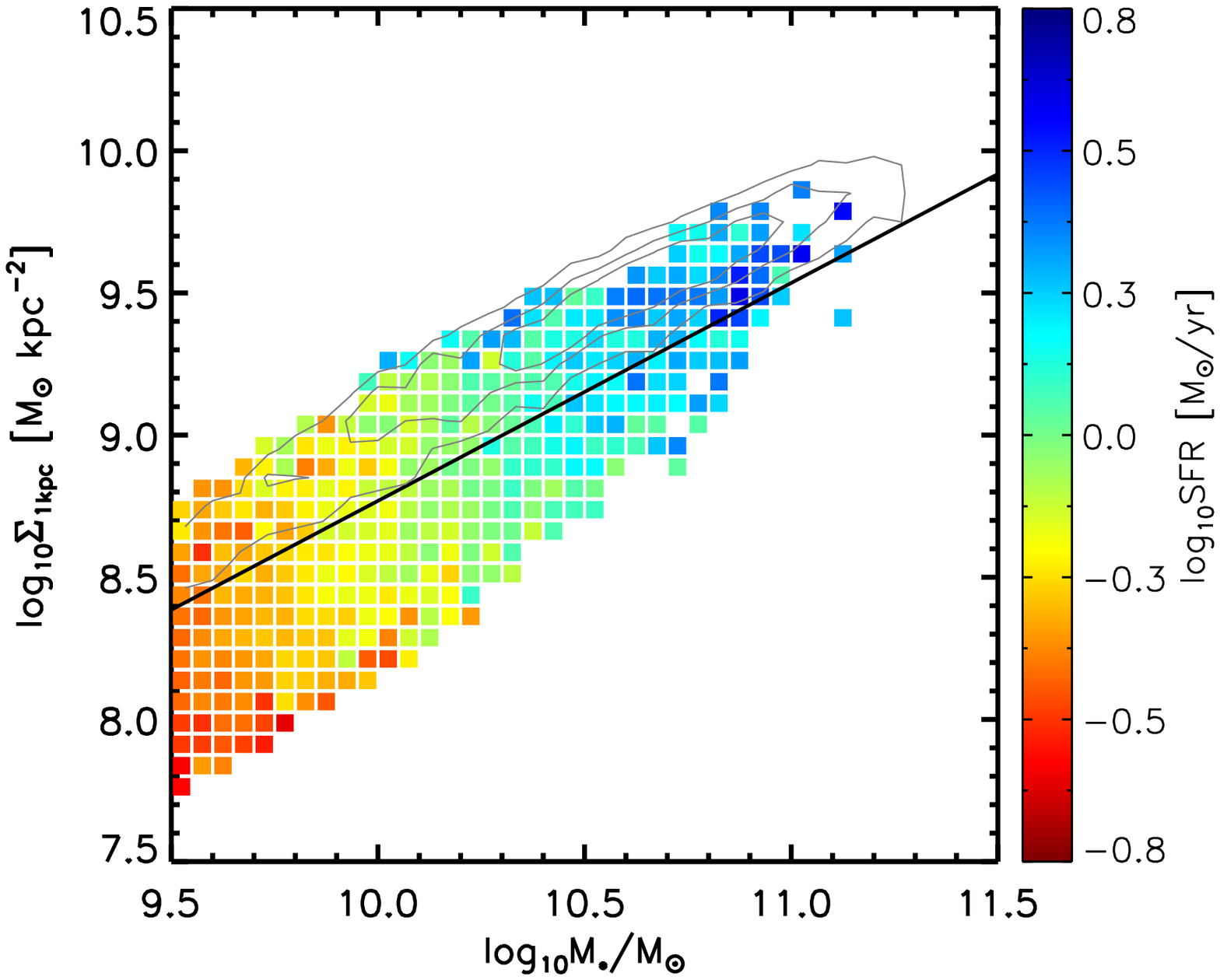,clip=true,width=0.405\textwidth}
   \epsfig{figure=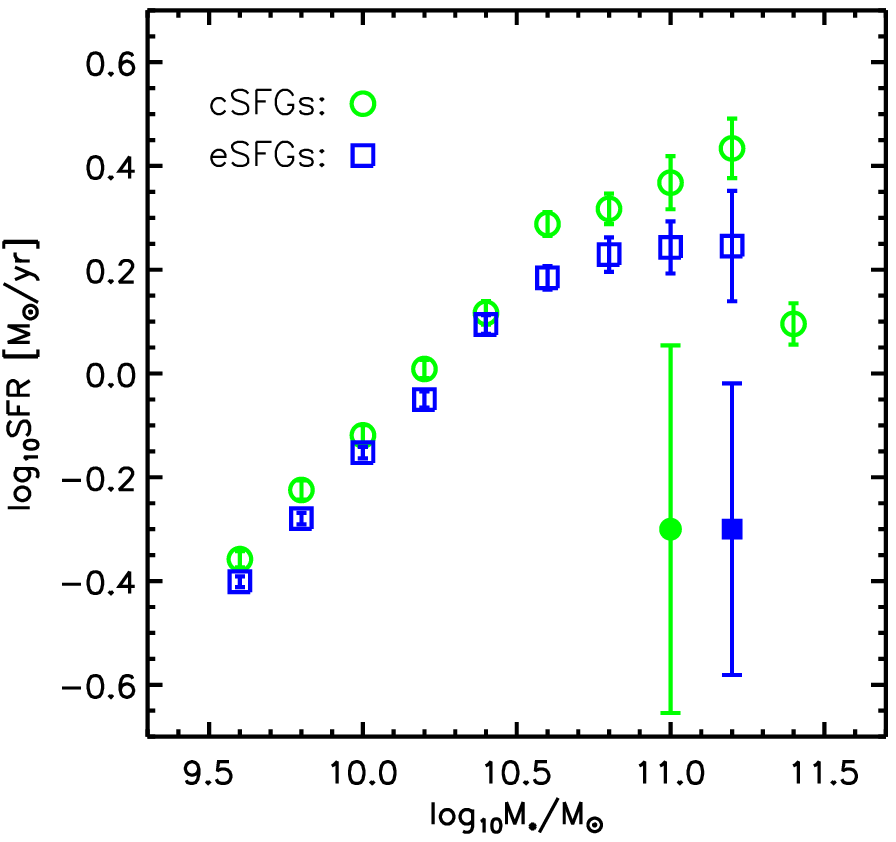,clip=true,width=0.378\textwidth}
   \epsfig{figure=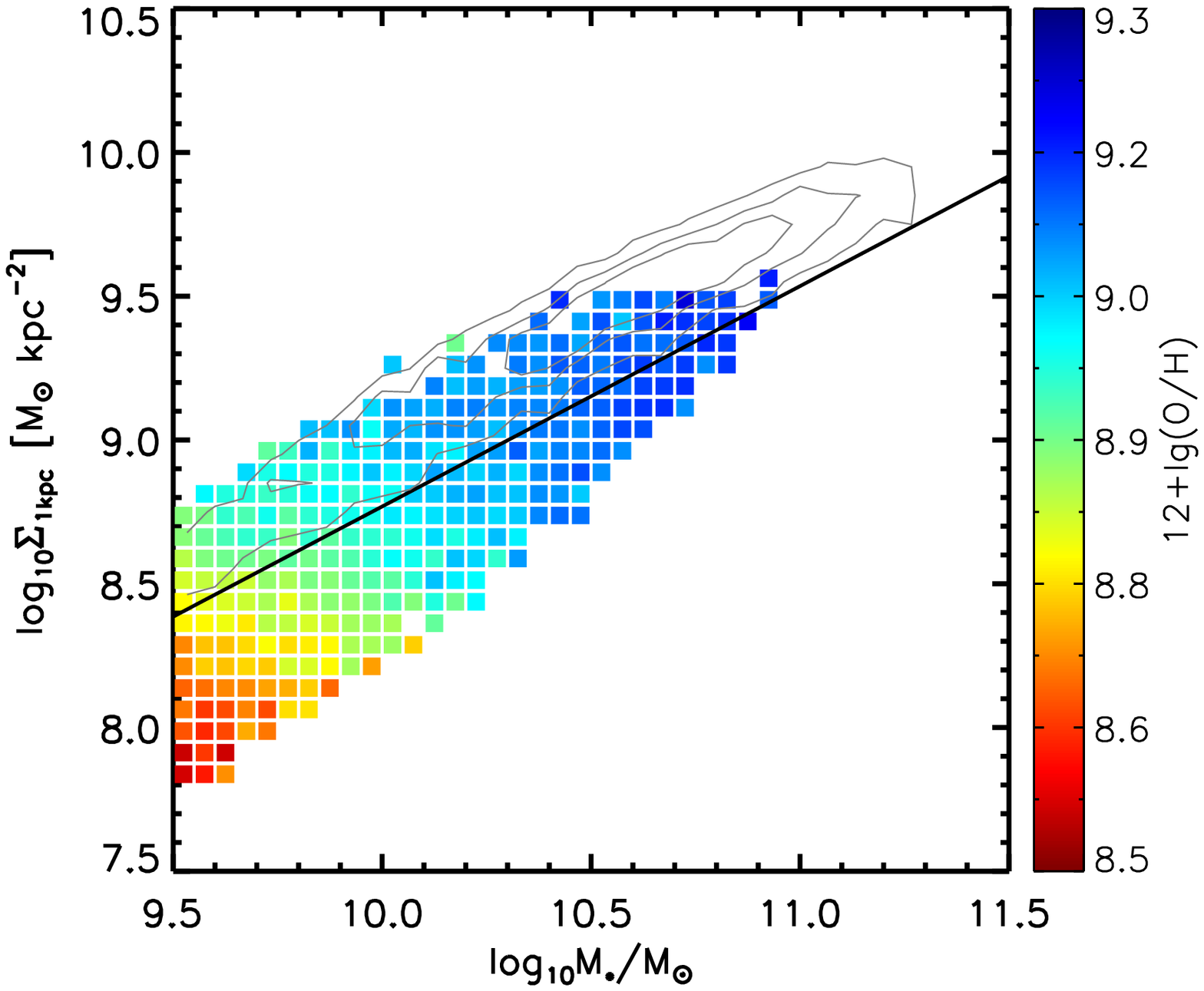,clip=true,width=0.405\textwidth}
  \epsfig{figure=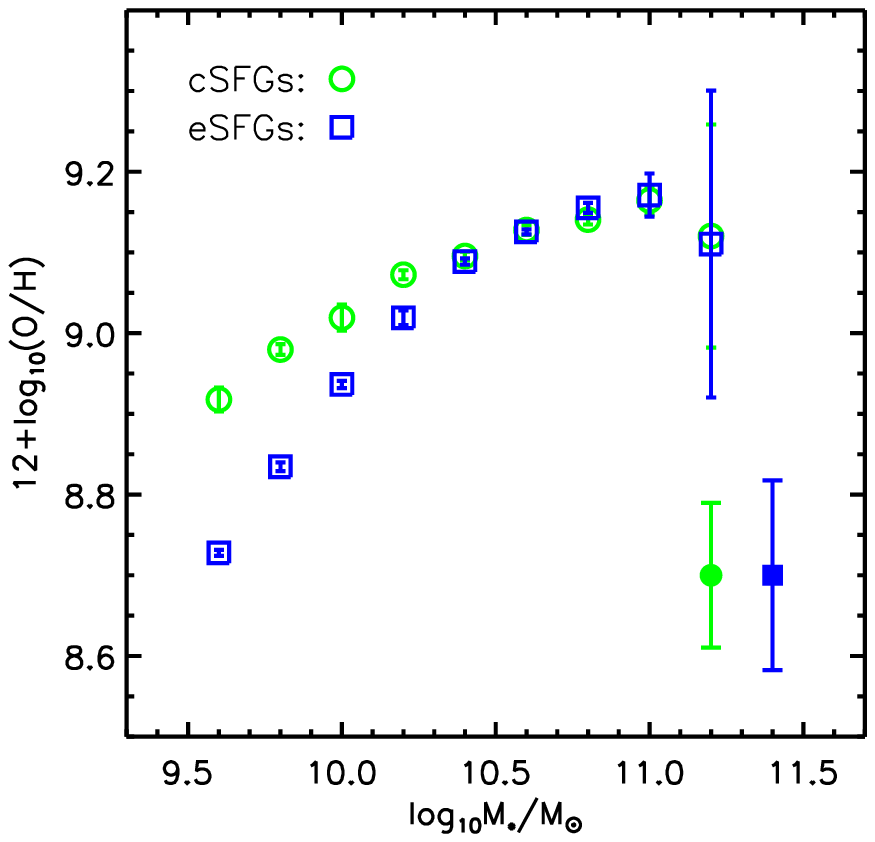,clip=true,width=0.378\textwidth}
\end{center}
\caption{The left two panels: the \sgm-\mstar\ diagram for SF galaxies color-coded with SFR (top) and gas-phase metallicity (bottom).   
The right two panels: the SFR (top) and gas-phase metallicity (bottom) as a function of stellar mass for cSFGs (green circles) and eSFGs (blue squares).  
In the left two panels, the contours and lines are the same as those in the left panel of Figure \ref{fig:size}. 
Similar to Figure \ref{fig:size}, the representative scatters of the relation for cSFGs and eSFGs are shown in the bottom-right corner of each right panel. 
}
\label{fig:sfr_zgas}
\end{figure*}

The SFR and gas-phase metallicity are two basic properties of SF galaxies. SFR reflects the present growth rate of stellar mass of galaxies, while metallicity plays an important role in many fundamental physical processes that regulate galaxy evolution, including gas cooling, star formation and dust formation. 
The tight SFMS established both in low and high redshift \citep[e.g.][]{Brinchmann-04, Salim-07, Daddi-07, Elbaz-11} suggests that the bulk of star formation in the universe occurs in a quasi-steady state \citep{Noeske-07b} and the fraction of the lifetime for a given SF galaxy during which it lies significantly above the SFMS is small. 
Similarly, the strong correlation between stellar mass and gas-phase metallicity has been found for decades \citep[e.g.][]{Lequeux-79, Tremonti-04, Ellison-08}. The drop of metallicity at low stellar mass end is usually interpreted as the evidence for the ubiquity of galactic winds and their feedback in removing metals from galaxy potential wells. Thus, investigating these two scaling relations for cSFGs and eSFGs would provide clues for compaction process under the scenario of compact and quenching. 

The left two panels of Figure \ref{fig:sfr_zgas} show the \sgm-\mstar\ relation for SF galaxies color-coded by SFR and gas-phase metallicity, respectively. 
At fixed stellar mass, the SFR appears to show no or very weak dependence on \sgm, suggesting that cSFGs and eSFGs appear to have similar SFR at fixed stellar mass as a whole, given their different structural properties. 
In contrast, the gas-phase metallicity is sensitive to \sgm\ at the low stellar mass end (\mstar$<10^{10.5}$\msolar), with the higher \sgm\ the higher gas-phase metallicity. This indicates that cSFGs are more metal-rich than eSFGs at low stellar mass end.   
We note that there is a lack of valid measurements of metallicity at high mass end (see the bottom-left panel of Figure \ref{fig:sfr_zgas}), which is due to the fact that the computing of metallicities requires galaxies to have 5$\sigma$ detection of the emission lines for H$\alpha$, H$\beta$, and [NII] \citep{Tremonti-04}.  

These results can be more clearly seen in the right two panels of Figure \ref{fig:sfr_zgas}, where the median SFR-\mstar\ and \zgas-\mstar\ relations are presented for cSFGs (green circles) and eSFGs (blue squares). As shown, the SFRs of cSFGs appear to be at most slightly higher than those of eSFGs (with an average of 0.06 dex), while we note that this result is preserved over the whole stellar mass range especially at high stellar mass end. This indicates that under the compaction and quenching scenario, galaxies sustain or even slightly enhance their star formation activities during the evolution from eSFGs to cSFGs. 
In contrast, for \zgas-\mstar\ relation, the pronounced differences between the two populations are shown. The differences become less significant with increasing stellar mass. Specifically, cSFGs are more metal-rich at \mstar$\sim10^{9.6}$\msolar\ than eSFGs for $\sim$0.19 dex, while the gas-phase metallicities of the two become comparable with each other at \mstar$\sim10^{10.5}$\msolar. This finding is consistent with the result of \cite{Ellison-08} that the gas-phase metallicity strongly depends on the galaxy half-light radius, in the sense that, more compact galaxies are more metal-rich at fixed stellar mass. Furthermore, the metallicity difference can be as large as 0.2 dex, which is roughly equal to what we find here. The dominated mechanism of this finding is still under debate \citep{Ellison-08, Yabe-12, SanchezAlmeida-14, Wang-17, SanchezAlmeida-DallaVecchia-18}, which is likely a combined effect of metal-poor gas accretion, metal-rich gas outflow, and the metal-enrichment from current star formation activities.  
However, the metallicity difference between cSFGs and eSFGs vanishes at stellar mass greater than $10^{10.5}$\msolar. This is likely due to the fact that the metallicities from MPA-JHU catalog are measured based on the 3-arcsec fiber spectra, which could not represent the global metallicity for large galaxies. Considering that SF galaxies with large stellar disk usually have significant metallicity gradient \citep[e.g.][]{Sanchez-14, Carton-18}, we suspect that the cSFGs likely have higher global metallicity than eSFGs even at high stellar mass if the metallicities are measured within the apertures that cover the whole galaxies \citep{Wang-17}.

Although the median SFR of the two populations are similar at fixed stellar mass, the distribution of SFR of the two can be different.  Indeed, we have checked the distribution of SFR for cSFGs and eSFGs at given stellar mass, and find that the SFR of cSFGs is more broadly distributed at high SFR end with respect to that of eSFGs. This is also confirmed by the standard deviation of SFR for cSFGs and eSFGs, shown in Figure \ref{fig:sfr_scatter}. We will discuss this in detail in Section \ref{subsec:depletion}.  

\subsection{The \HI\ properties}
\label{subsec:HI}

\begin{figure*}
 \begin{center}
   \epsfig{figure=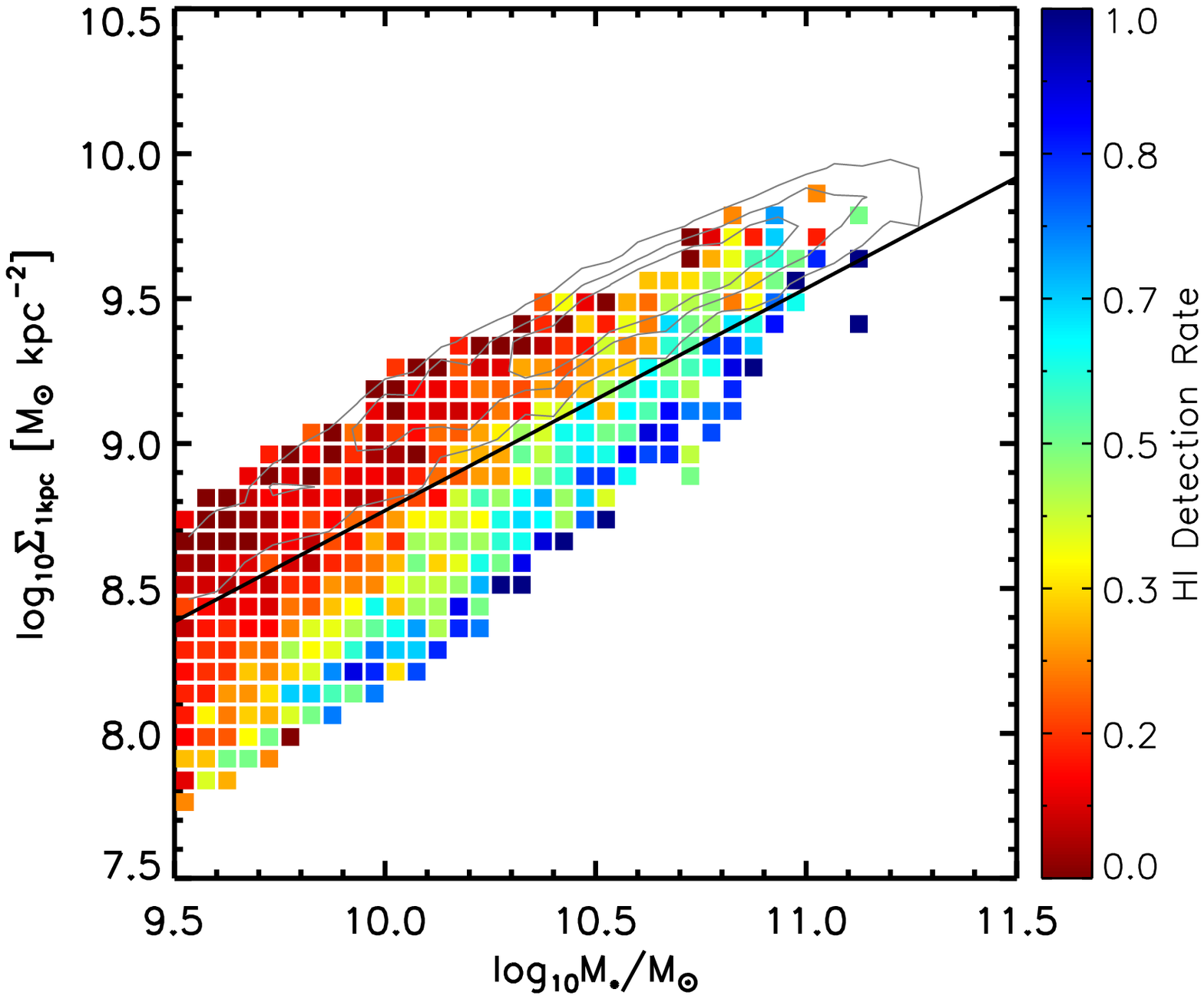,clip=true,width=0.405\textwidth}
   \epsfig{figure=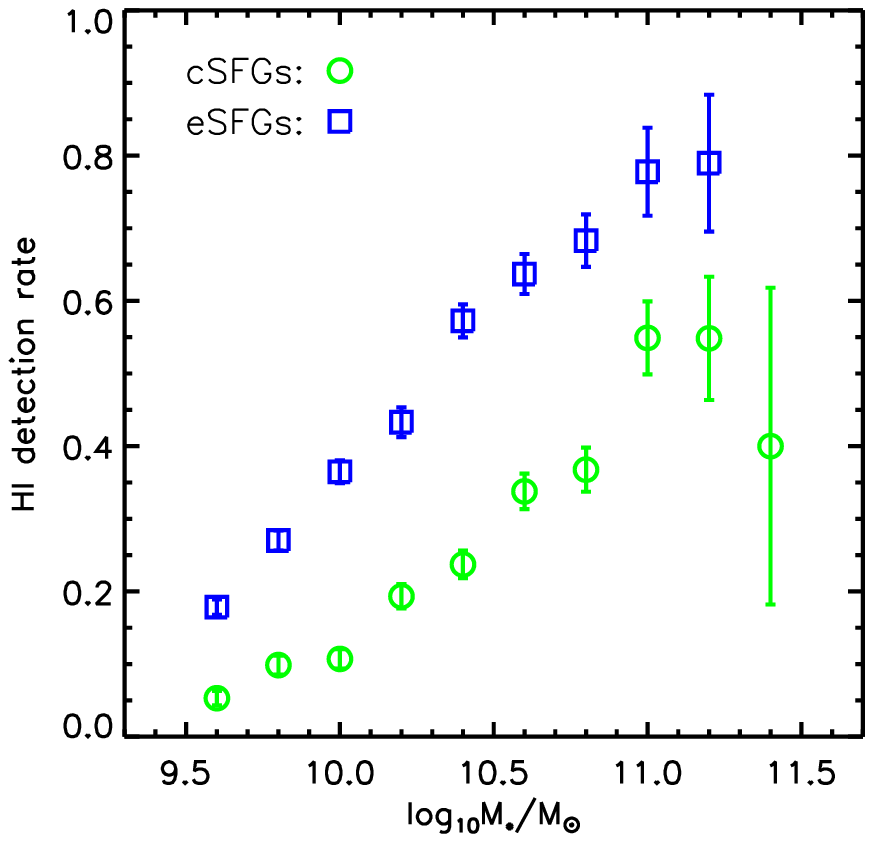,clip=true,width=0.378\textwidth}
   \epsfig{figure=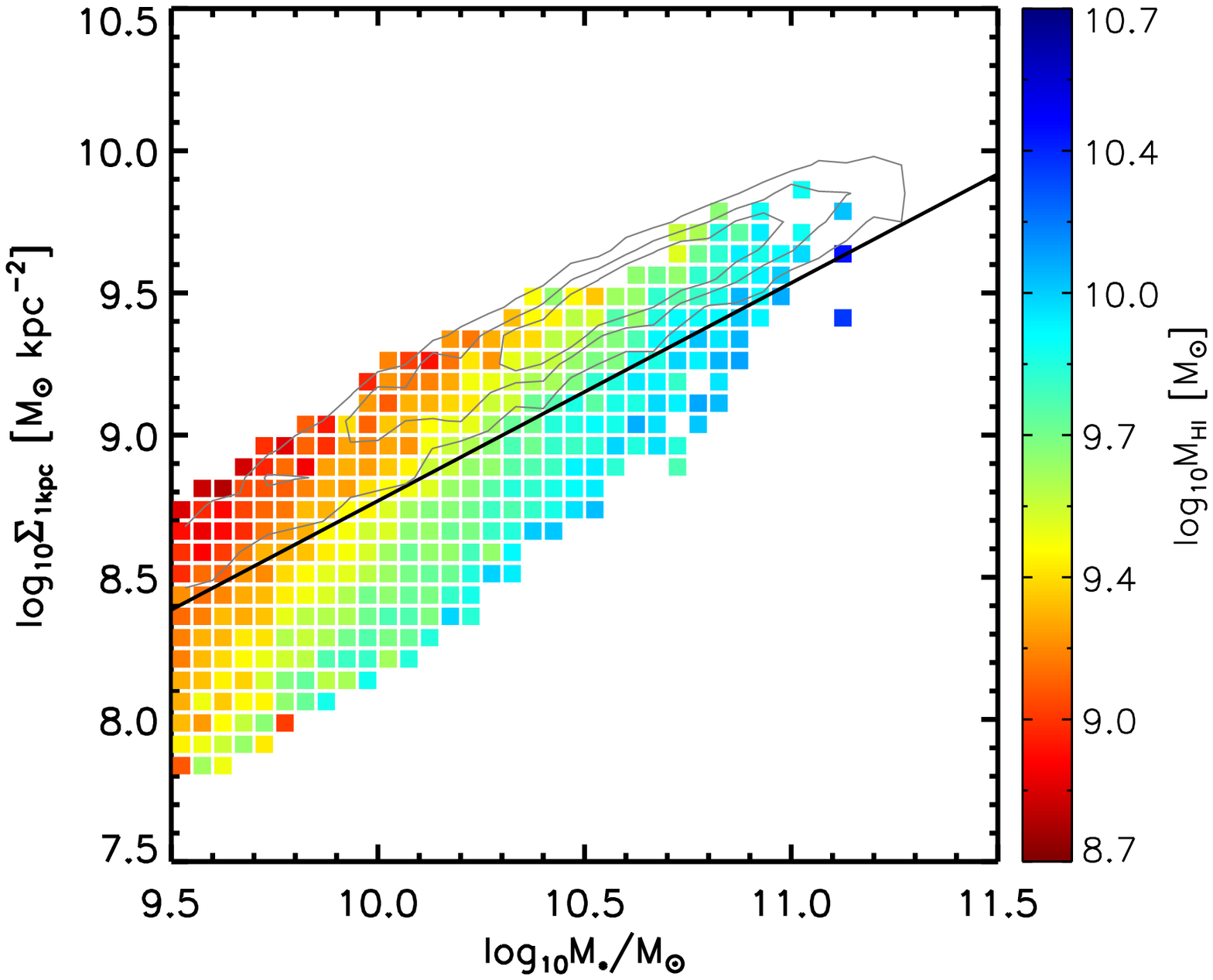,clip=true,width=0.405\textwidth}
  \epsfig{figure=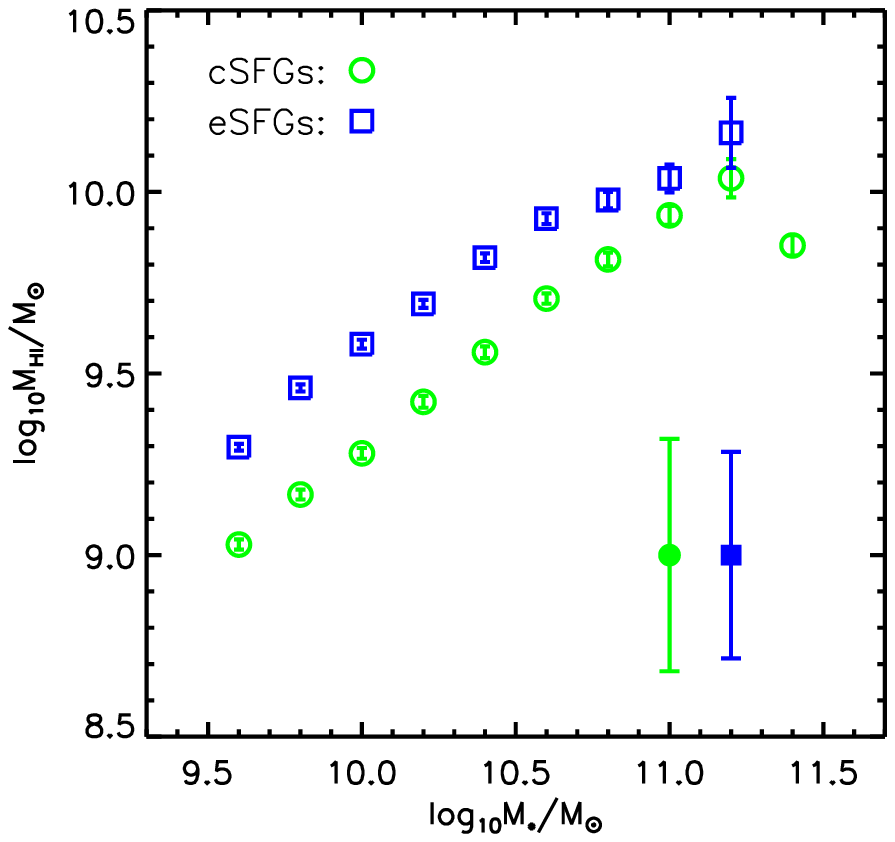,clip=true,width=0.378\textwidth}
\end{center}
\caption{The same as Figure \ref{fig:sfr_zgas}, but for \HI\ detection rate and \HI\ mass. Lines, colors and symbols are the same as those in Figure \ref{fig:sfr_zgas}. }
\label{fig:map_HI}
\end{figure*}

The cold gas in galaxies is the fuel to sustain their star formation \citep[e.g.][]{Sancisi-08, Conselice-13, Wang-13, Wang-15}. Galaxies are usually quenched by exhausting or heating their cold gas, removing their cold gas away, or stabilizing the gas disk against gravitational instabilities to suppress star formation. Thus, investigating the cold gas properties in galaxies is one of the most important keys to uncover the process of star formation quenching. 
For instance, by using \HI\ spectral stacking of a sample of early-type galaxies in ALFALFA footprint, \cite{Fabello-11} did not find the evidence that galaxies with a prominent bulge component are less efficient in turning their cold gas reservoirs into stars, which appears to be inconsistent with the ``morphological quenching'' picture \citep{Martig-09}. 
In this subsection, we will examine the \HI\ properties of cSFGs and eSFGs, and the possible evolution of gas content during the compaction process under the compaction and quenching scenario.

Since just a small fraction of the sample galaxies are matched with ALFALFA galaxies, we could not obtain the \HI\ properties for each of the sample galaxies. 
This means that direct comparison of the observed \HI\ content for all the cSFGs and eSFGs is not valid.
Alternatively, we introduce a parameter, the \HI\ detection rate, defined as the fraction of galaxies that are detected by ALFALFA survey for a given subsample. Thus, for a given subsample of fixed stellar mass, the higher \HI\ detection rate is usually corresponding to the higher averaged \HI\ content, since ALFALFA is a blind \HI\ survey. Due to the detection limit of ALFALFA, there is a lack of \HI\ sources with the \HI\ mass below $10^{9.7}$\msolar\ at redshift of 0.05, indicating that the detection limit of \HI\ gas is $\sim10^{9.7}$\msolar\ by this redshift.  Thus, in order to reduce the effect of the detection limit, galaxies only with \HI\ mass greater than $10^{9.7}$\msolar\ are treated as \HI\ detected sources. In this way, there are in total 19.1\% of cSFGs and 35.8\% of eSFGs treated as HI detected sources. 

The top left panel of Figure \ref{fig:map_HI} shows the \sgm-\mstar\ diagram with the color-coding of the \HI\ detection rate. At fixed \sgm, the \HI\ detection rate slightly increases with increasing stellar mass, which is consistent with the fact that more massive SF galaxies usually host more \HI\ gas \citep{Catinella-10, Wang-15}. However, at fixed stellar mass, the \HI\ detection rate shows rapid decline with increasing \sgm. 
This result can be clearly seen in the top right panel of Figure \ref{fig:map_HI}, where the \HI\ detection rate as a function of stellar for cSFGs and eSFGs are presented. As shown, the eSFGs exhibit higher \HI\ detection rate with respect to cSFGs, suggesting that eSFGs are on average more gas-rich than cSFGs. 
The bottom left panel of Figure \ref{fig:map_HI} presents the \sgm-\mstar\ diagram color-coded with the \HI\ mass.  The \HI\ masses for galaxies with undetected 21 cm emission in ALFALFA are estimated with adopting the relation (see Section \ref{subsec:parameters}) from \cite{Catinella-10}. 
As shown,  the \HI\ mass is strongly anti-correlated with \sgm, indicating that eSFGs tend to be more \HI-rich than cSFGs. Indeed, as shown in the bottom right panel of Figure \ref{fig:map_HI}, the eSFGs have higher median \HI\ mass than cSFGs for $\sim$0.3 dex, which confirms the result found in top two panels of Figure \ref{fig:map_HI}. 

We conclude that eSFGs are more \HI-rich than cSFGs, which is consistent with the compaction and quenching scenario with assuming that galaxies evolve along the direction of cold gas consumption. 

\subsection{The Environment of cSFGs and eSFGs}
\label{subsec:environment}

\begin{figure*}
 \begin{center}
   \epsfig{figure=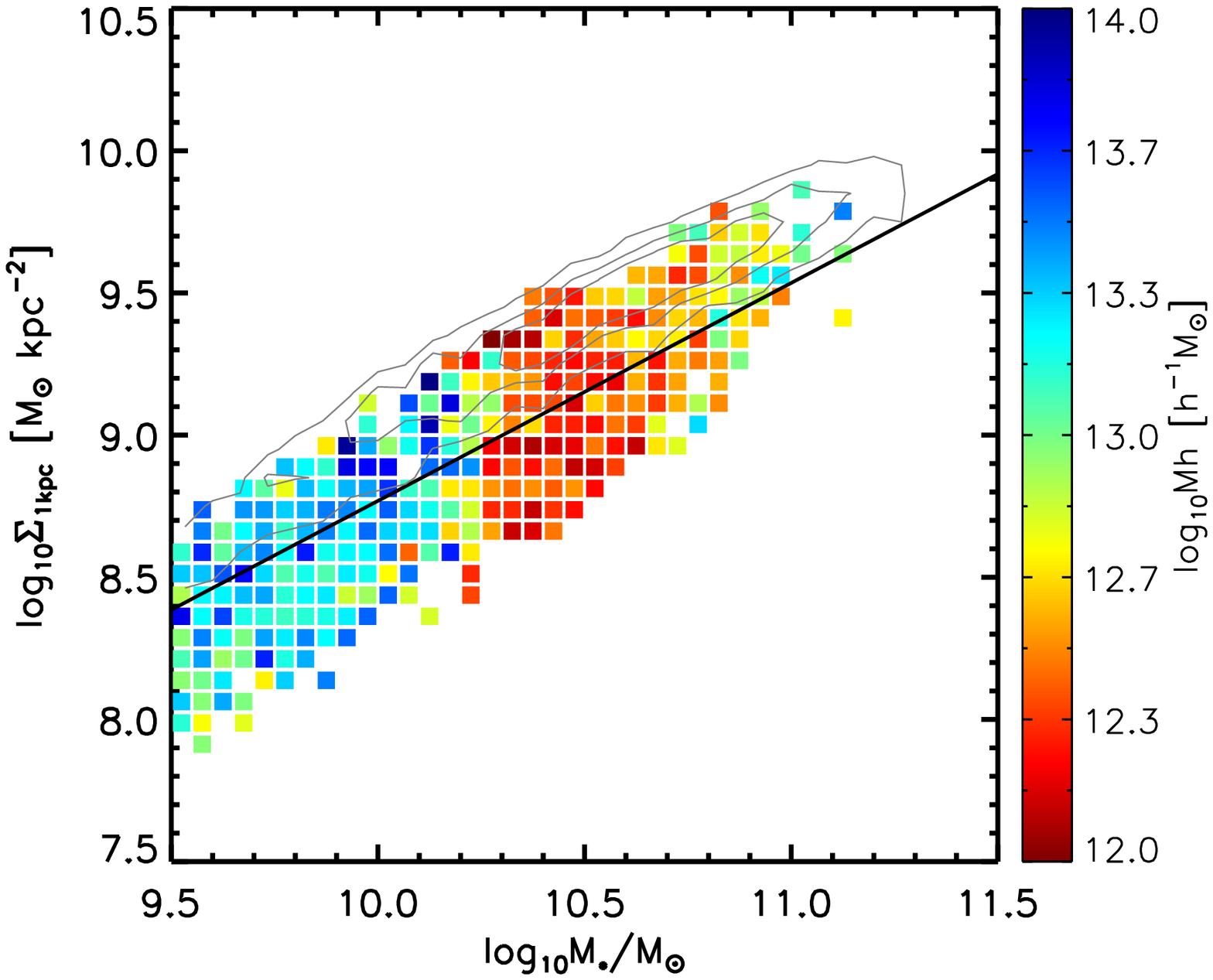,clip=true,width=0.405\textwidth}
   \epsfig{figure=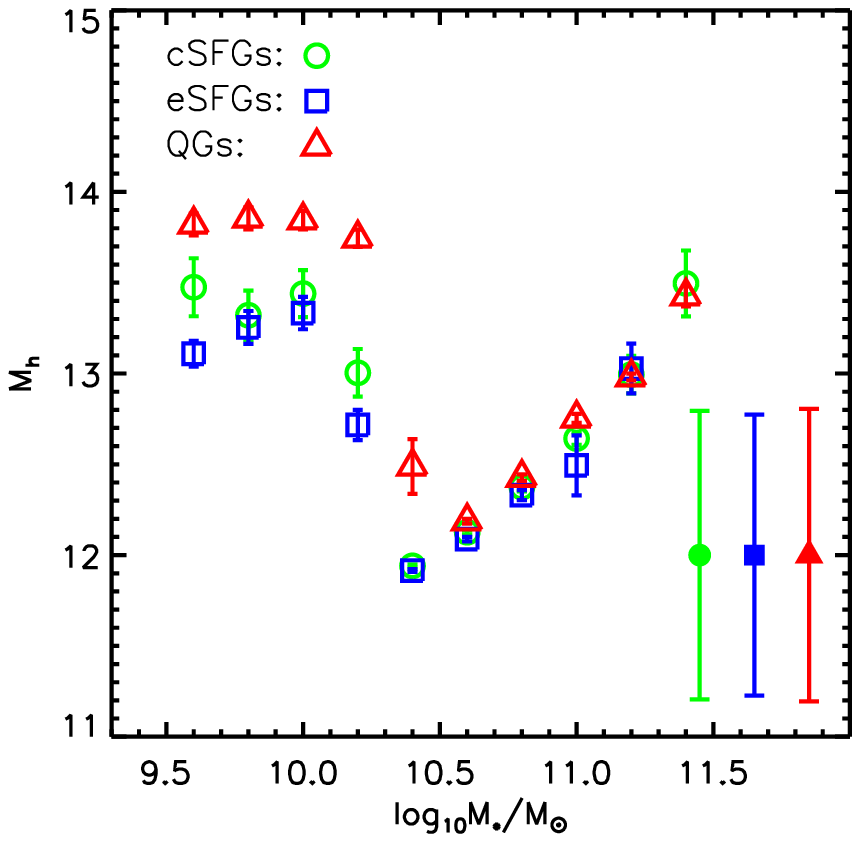,clip=true,width=0.378\textwidth}
   \epsfig{figure=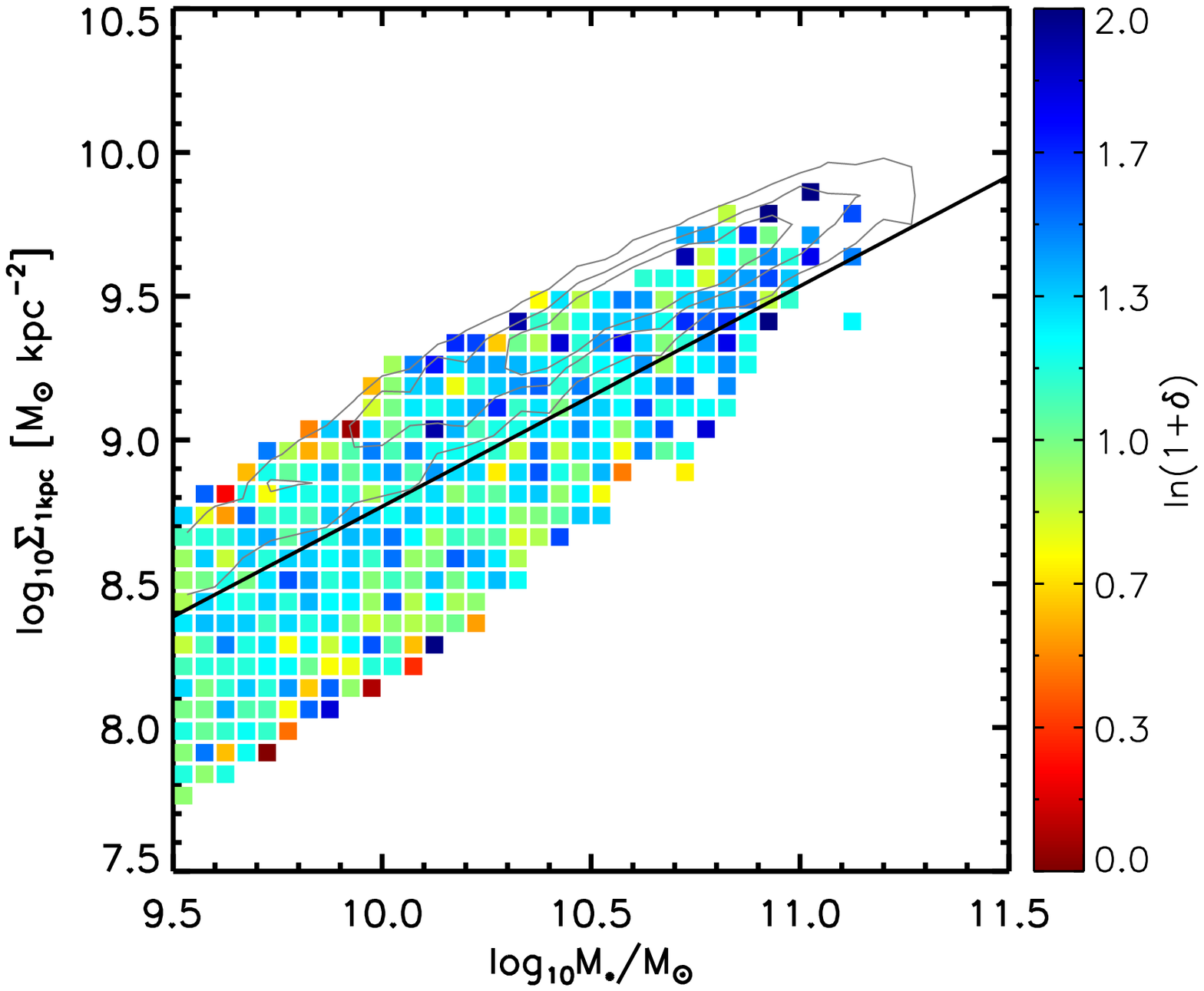,clip=true,width=0.405\textwidth}
  \epsfig{figure=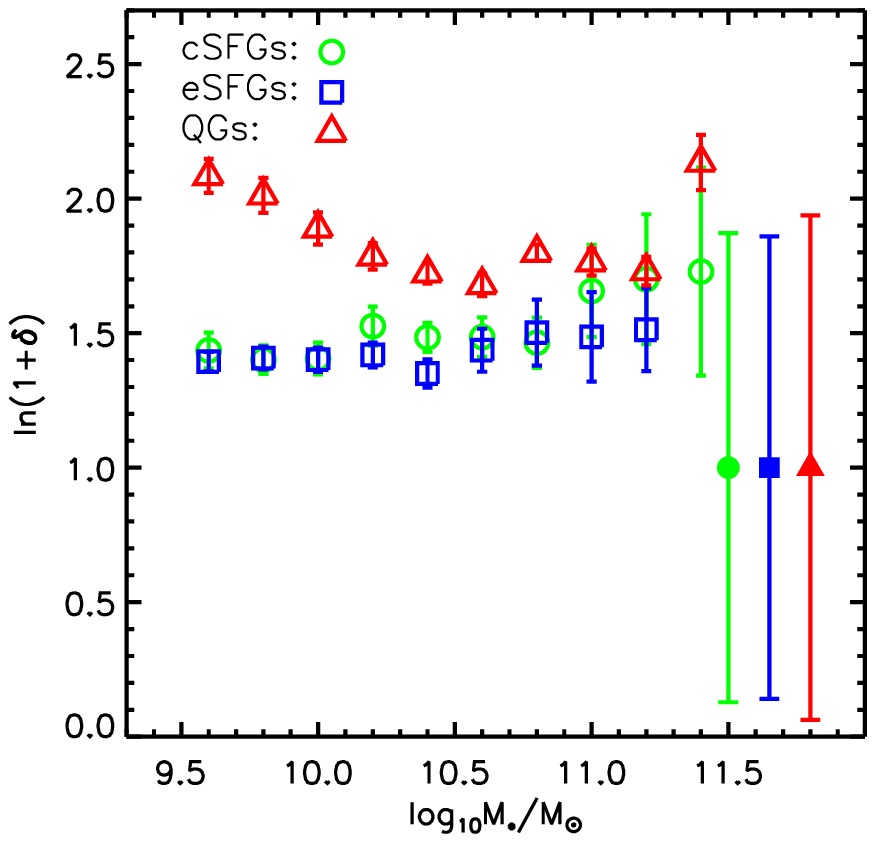,clip=true,width=0.378\textwidth}
\end{center}
\caption{The same as Figure \ref{fig:size}, but for halo mass and overdensity. Lines, colors and symbols are the same as those in Figure \ref{fig:size}. }
\label{fig:environ}
\end{figure*}

Galaxies resided in different environments have different prevalence of quenched population \citep[e.g.][]{Peng-10, Woo-15, Wang-18c}, indicating that environmental effects play an important role in quenching a galaxy.
The fraction of quenched galaxies is found to be correlated with halo mass, and anti-correlated with halo-centric radius \citep[e.g.][]{Weinmann-06, vandenBosch-08, Wetzel-Tinker-Conroy-12, Kauffmann-13, Wang-18b, Wang-18c}, indicating that galaxies resided in the inner region of massive halos are more likely to be quenched with respect to galaxies in the outer region of the similar halos or the inner region of less massive halos. Not only working on the star formation quenching, environmental effects are believed to effectively transform the morphology of galaxies from disk-like to spheroid-like. For instance, major mergers can efficiently transform the SF disk galaxies to quenched ellipticals. In addition, numerical simulations have shown that disk galaxies in a rich cluster would suffer from the frequent encounters with member galaxies and the cluster's tidal field (also known as galaxy harassment), along with a complete morphological transformation from disks to spheroids \citep[e.g.][]{Moore-96, Moore-Lake-Katz-98, Fujita-98}. 
However, it is still unclear what the role of environment is in compaction process. In this subsection, we will examine environment of cSFGs and eSFGs, and try to find whether the environment plays a key role in compaction process under the scenario of compaction and quenching. 

The left two panels of Figure \ref{fig:environ} show the \sgm-\mstar\ relation for SF galaxies with the color-coding of the host halo mass and the local overdensity. As shown, the host halo mass of both cSFGs and eSFGs strongly depends on stellar mass, while the overdensity of the two populations shows no or very weak dependence on the stellar mass. Specifically, the halo mass of SF galaxies shows a dramatically drop at the stellar mass of $\sim10^{10.3}$\msolar. This is because satellite galaxies dominate the galaxy population at low stellar mass end, while central galaxies dominate the population at high stellar mass end \citep{Yang-Mo-vandenBosch-09}. 
In contrast, the overdensity of SF galaxies is almost independent of stellar mass, which seems to be inconsistent with the dependence of halo mass on stellar mass. Actually, this discrepancy is due to the fact that the SF satellite galaxies of low stellar mass tend to reside in the outer region of host halos, which makes their local overdensity to be comparable with those of high-mass central galaxies. 
In the present work, we do not divide galaxies into centrals and satellites, because centrals and satellites are found to exhibit similar quenching behaviors when both halo mass and stellar mass are controlled \citep{Hirschmann-14,  Knobel-15, Wang-18b, Wang-18c}. 
More importantly, at fixed stellar mass, the halo mass and/or overdensity of SF galaxies exhibit no or very weak dependence on \sgm, indicating that cSFGs and eSFGs reside in halos of similar mass with similar overdensities. This result can be more clearly seen in the right panels of Figure \ref{fig:environ}, where the median halo mass and overdensity as a function of stellar mass for cSFGs, eSFGs and quenched galaxies are shown. We find that the quenched galaxies more likely reside in more massive halos (or regions with higher matter densities) with respect to cSFGs or eSFGs at low stellar mass end (\mstar$<10^{10.5}$\msolar), suggesting an environmental driven quenching mechanism especially for less massive galaxies.   
However, we find the environments of cSFGs and eSFGs do not shows significant differences, suggesting that environmental effects do not play a major role in compaction process under the scenario of compaction and quenching at least for low-redshift universe.  

\section{Discussion}
\label{sec:discussion}


\subsection{Gas depletion time along the SFMS}
\label{subsec:depletion}

\begin{figure*}
  \begin{center}
   \epsfig{figure=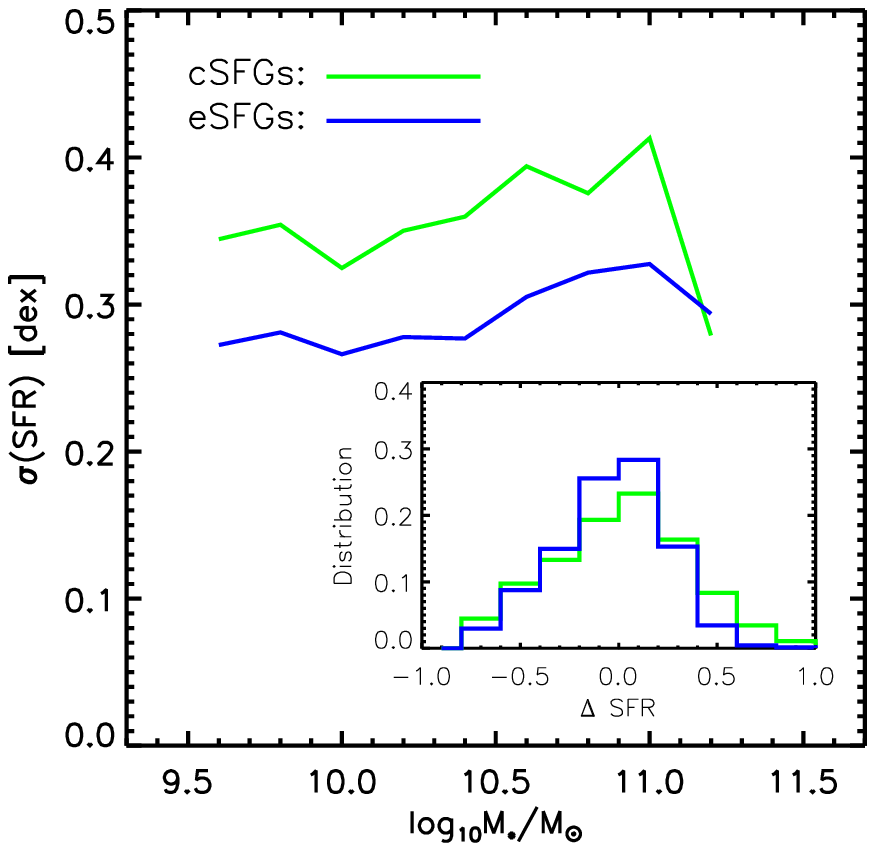,clip=true,width=0.378\textwidth}
   \epsfig{figure=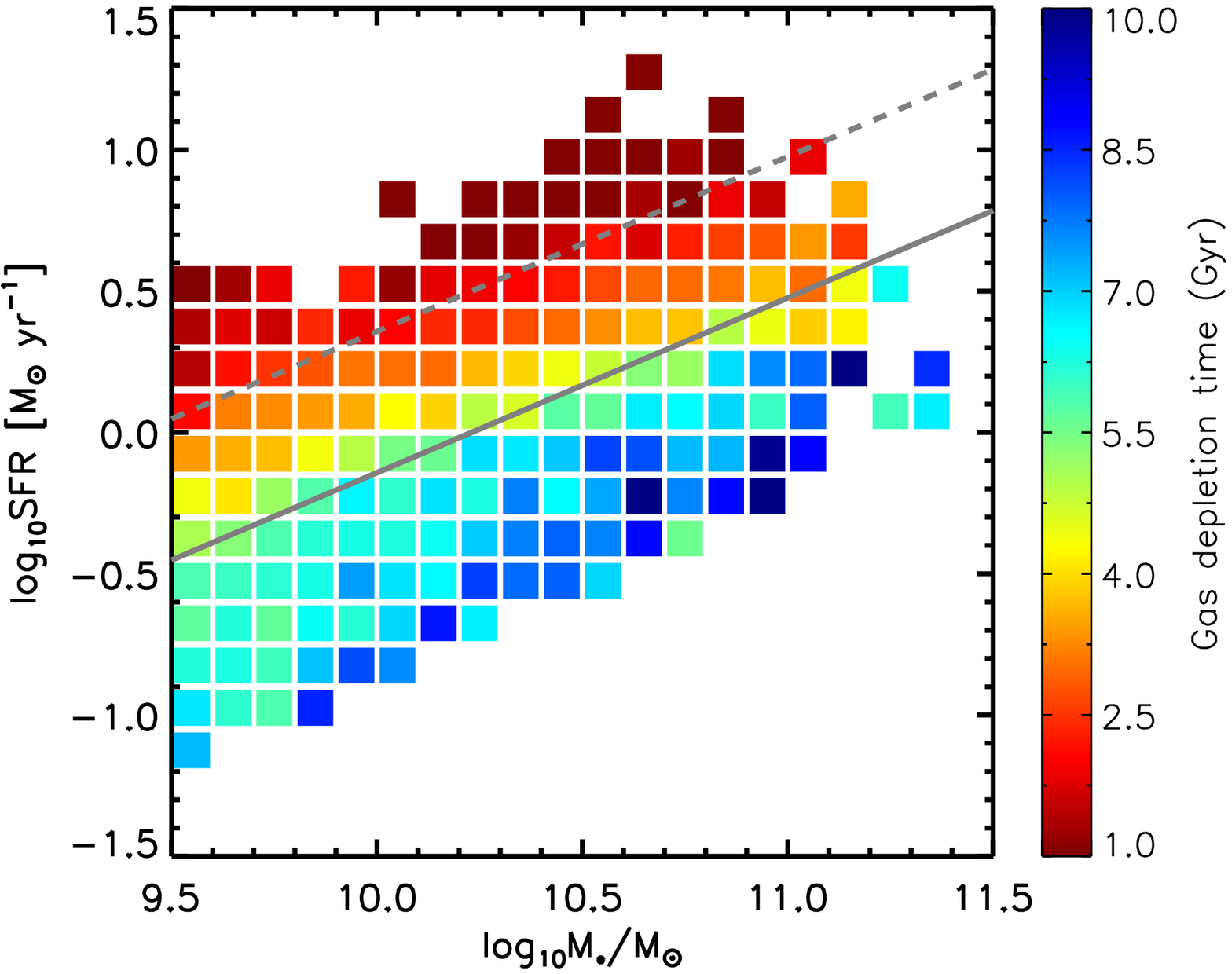,clip=true,width=0.405\textwidth}
  \end{center}
\caption{Left panel: The standard deviation of SFR as a function of stellar mass for cSFGs (green) and eSFGs (blue).   The normalized distribution of $\Delta$SFR for cSFGs (green histogram) and eSFGs (blue histogram) are also presented in the small panel inside. 
Right panel: the SFR-\mstar\ relation for SF galaxies with the color-coding of gas depletion time. The solid gray line is the best-fit relation of SFMS taken from Figure \ref{fig:sample_definition}, and the dashed gray line is the 0.5 dex above this relation. }
 \label{fig:sfr_scatter}
\end{figure*}

In Section \ref{subsec:sfr}, we find that cSFGs appear to have slightly higher SFR than eSFGs. However, the SFR distribution of cSFGs and eSFGs can be different. 
We present the standard deviation of the SFR as a function of stellar mass for cSFGs and eSFGs in the left panel of Figure \ref{fig:sfr_scatter}. 
The scatter of SFMS shown here is $\sim$0.3 dex, which is broadly consistent with the previous findings of 0.2-0.3 dex \citep[e.g.][]{Noeske-07b,Whitaker-12, Speagle-14}. 
As one can see, the standard deviations of SFR for cSFGs are systematically higher than those of eSFGs almost over\ the whole stellar mass range, suggesting that cSFGs have wider SFR distribution than eSFGs at fixed stellar mass. 
Indeed, we check the distribution of the two population on the SFR-\mstar\ diagram and find that 288 ($\sim8.3$\%) cSFGs lie above the normal SFMS for 0.5 dex, while only 86 ($\sim1.7$\%) eSFGs lie above the normal SFMS for 0.5 dex, indicating that galaxies with the most intense star formation activities tend to be cSFGs, rather than eSFGs at given stellar mass (see Figure \ref{fig:sample_definition}). This result is confirmed by the normalized distribution of $\Delta$SFR, defined as the deviation from the normal SFMS in logarithmic space, for eSFGs and cSFGs.
This result is also in good agreement with the cosmological simulations of \cite{Tacchella-16a}, who found that galaxies in the upper envelope of the SFMS tend to be compact at redshift of $>1$.
We have checked the morphologies of these cSFGs with $\Delta$SFR$>$0.5 dex by visually inspection, and find that only a few of them ($<$10\%) exhibit clear features of on-going major merger. 

In the right panel of Figure \ref{fig:sfr_scatter}, we present the SFR-\mstar\ relation for SF galaxies with the color-coding of \HI\ depletion time. 
The \HI\ depletion time\footnote{The \HI\ masses of \HI\ non-detected galaxies in ALFALFA are estimated using the relation from \cite{Catinella-10}.} is computed with assuming that 1) galaxies sustain their current SFR, and 2) the replenishment of cold gas is negligible \citep{Tacchella-16a}.  
We do not consider the contribution of Helium and the molecular gas in the calculation. 
As shown, the \HI\ depletion time for galaxies with most intense star formation activities can be as low as 1 Gyr. These galaxies are dominated with cSFGs, and have strong star formation activities, suggesting that they likely suffer from the on-going compaction along with the rapid build-up of their stellar cores (or bulges) and the rapid cold gas consumption. 

In addition, we examine the median gas depletion time for cSFGs and eSFGs with the above assumptions. We find that the \HI\ depletion time only weakly depends on stellar mass, with the median \HI\ depletion time of $\sim$4 Gyr for cSFGs, and $\sim$8 Gyr for eSFGs. This suggests that cSFGs are statistically more close to quenched population than eSFGs along the evolution track. 
However, in computing \HI\ depletion time, we assume that SF galaxies sustain their current SFR, which is not true in the real universe. Indeed, by investigating the simulated SF galaxies of high redshift ($z>$1), \cite{Tacchella-16a} proposed that SF galaxies go up and down across the normal SFMS before quenching their star formation. Observationally, we find a small but significant fraction of cSFGs that lies above the normal SFMS for 0.5 dex. This suggests that the fraction of the lifetime for a cSFG during which it lies significantly above the normal SFMS is small but significant, assuming that lying significantly above the normal SFMS is a period of a SF galaxy during its lifetime. If this is the case, the cSFGs consume a bulk of their cold gas and build their stellar cores in a short timescale when they lie significantly above the normal SFMS, subsequently they are back to the normal SFMS or even lower. Thus, it is not necessary to take 4-8 Gyr to quench a galaxy, because a bulk of its cold gas can be consumed in a short time scale when it lie significantly above the normal SFMS. 

Assuming galaxies evolve from eSFGs, through cSFGs to quenched population, the relative number of eSFGs and cSFGs would give instructions on the timescale of compaction and quenching. 
The fraction of cSFGs takes the proportion of $\sim$21.6\% of the whole sample, and $\sim$40.9\% of all the SF galaxies, indicating that cSFGs are a non-negligible galaxy population. This suggests that the star formation quenching is likely not dominated by rapid processes at local universe. This is different from the case of high-redshift, where major mergers and cold gas accretions are believed to much more frequently occur than in local universe \citep[e.g.][]{Rodriguez-Gomez-15, Tacchella-16a, Tacchella-16b}, suggesting a different compaction and quenching timescale with respect to the low-redshift universe. 

\subsection{The prevalence of AGN activities}

\begin{figure}
  \begin{center}
   \epsfig{figure=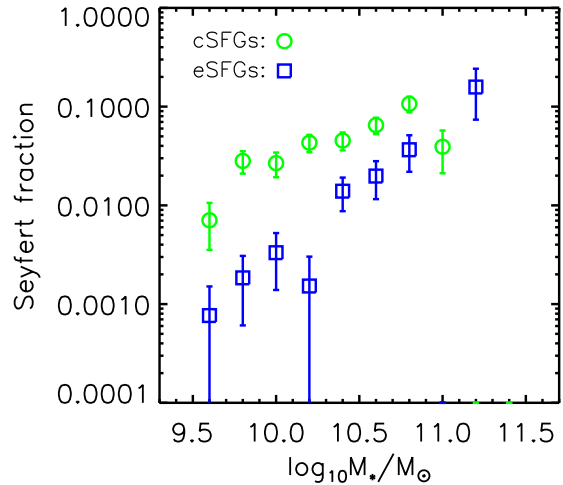,clip=true,width=0.45\textwidth}
  \end{center}
\caption{The Seyfert galaxy fraction as a function of stellar mass for eSFGs (blue squares) and cSFGs (green circles). }
 \label{fig:agn_activity}
\end{figure}

The AGN activities have been brought in to explain the link between inner structural properties of galaxies and star formation quenching \citep[e.g.][]{Croton-06, Fabian-12, Cicone-14, Henriques-15}, acting to expel/deplete the pre-existing cold gas supply, and/or prevent gas cooling and further star formation by strong feedback from central massive black holes. 
Since the build-up of stellar cores of cSFGs is probably connected to the growth of central massive black holes, the prevalence of AGN in cSFGs and eSFGs are likely to be different. Indeed, by examining the fraction of massive cSFGs that host an AGN at redshift of $\sim$2, \cite{Kocevski-17} found that $\sim$40\% of compact SF galaxies host an X-ray detected AGN, which is 3.2 times higher than the incidence of AGN in extended SF galaxies of similar masses and similar redshifts, based on the combination of Hubble WFC3 imaging and Chandra X-ray observations. In low-redshift universe, \cite{Wang-17} have identified a sample of galaxies with ``outside-in'' assembly mode from MaNGA \citep[Mapping Nearby Galaxies at Apache Point Observatory;][]{Bundy-15}, which are selected to have lower 4000\AA\ break in the inner region than in the outer region. They found that these galaxies are likely in the transitional phase from normal SF galaxies to quenched galaxies, given their smaller size, higher SFR and higher gas-phase metallicity with respect to normal SF galaxies of similar stellar mass. The properties of these galaxies resemble the properties of cSFGs investigated in this work, although the two populations are defined in totally different ways. 
In addition, the galaxies with outside-in assembly mode are more likely to host an AGN than normal SF galaxies, although with the limited sample size \citep{Wang-17, Liu-18}, suggesting a more prevalence of AGN in cSFGs than eSFGs. 

Figure \ref{fig:agn_activity} shows the fraction optical AGNs for cSFGs and eSFGs as a function of stellar mass. The optical AGNs are identified on the basis of the BPT diagram \citep{Baldwin-Phillips-Terlevich-81}, with adopting the division lines of Seyfert galaxies from \cite{Kewley-01} and \cite{CidFernandes-10}. The relevant emission lines are drawn from MPA-JHU catalog, with restricting the signal-to-noise ratios to be greater than 3.0. 
The emission line fluxes are corrected for intrinsic extinction based on the Balmer decrement with assuming a dust attenuation curve of the form $\lambda^{-0.7}$ \citep{Charlot-Fall-00} and an intrinsic flux ratio H$\alpha/$H$\beta=2.86$.
As shown in Figure \ref{fig:agn_activity}, the fraction of Seyfert galaxies is increasing with increasing stellar mass for both cSFGs and eSFGs. More importantly, the cSFGs are more likely to host an AGN than eSFGs almost over the whole stellar mass range, and the differences between the two population are more pronounced for low mass galaxies. Our result confirms the result that cSFGs are more likely to host an AGNs than eSFGs. However, that whether the more prevalence of AGN in cSFGs than in eSFGs is the by-product of compaction process, or the cause of star formation quenching is still under debate \citep[e.g.][]{Lilly-Carollo-16, Kocevski-17, Wang-18a}. 

\subsection{Implications of the result}
\label{subsec:implication}

\begin{figure*}
  \begin{center}
   \epsfig{figure=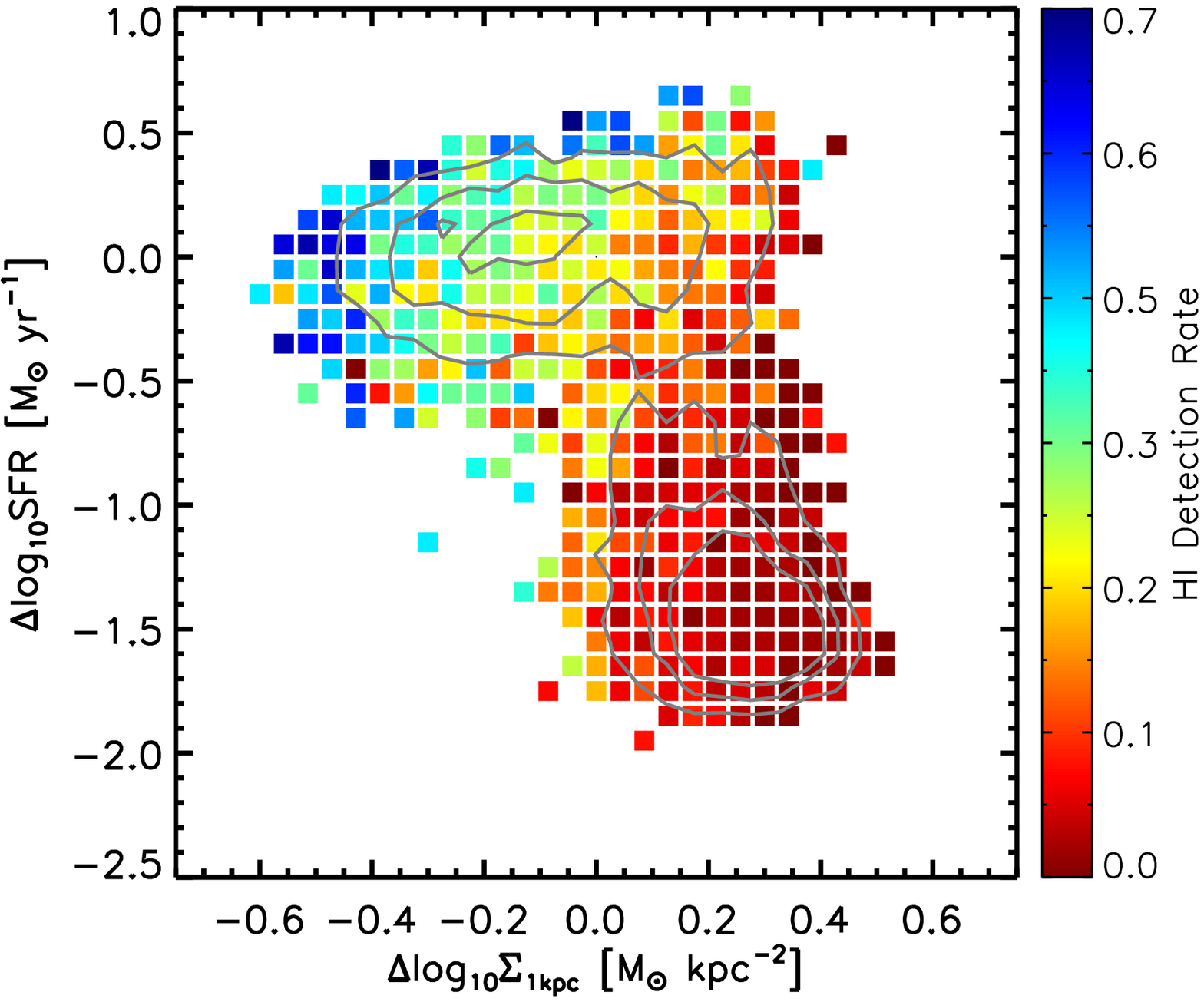,clip=true,width=0.45\textwidth}
   \epsfig{figure=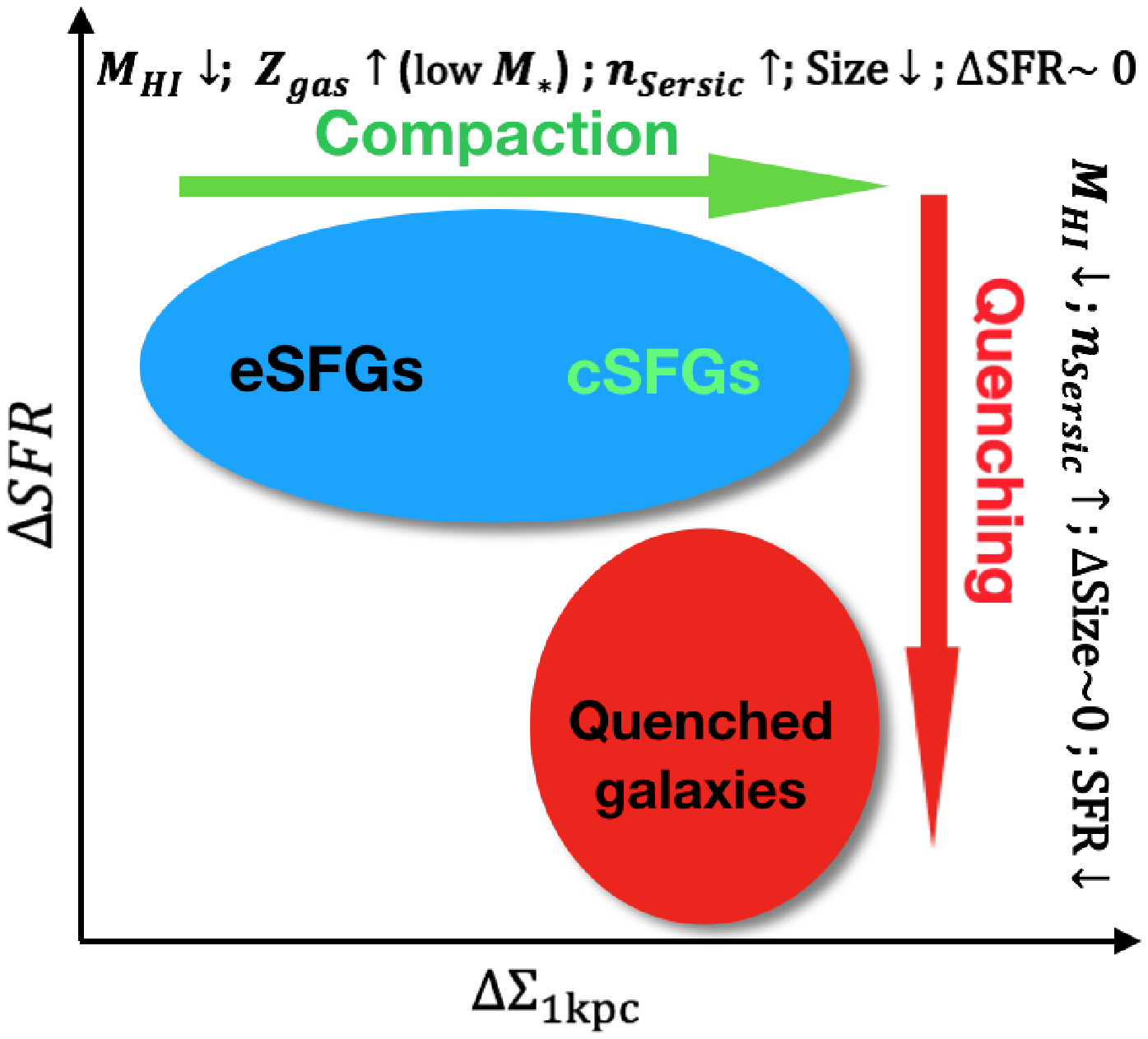,clip=true,width=0.41\textwidth}
   \end{center}
\caption{Left panel: The $\Delta$SFR-$\Delta\Sigma_{\rm 1kpc}$ diagram for all the sample galaxies color-coded with \HI\ detection rate. The gray contours shows the distribution of galaxies on this diagram.  Right panel: The sketch for galaxy evolution on the $\Delta$SFR-$\Delta\Sigma_{\rm 1kpc}$ diagram under the scenario of compaction and quenching.}
 \label{fig:HI_diagram}
\end{figure*}

As shown in Figure \ref{fig:sample_definition}, we present a simple line on the \sgm-\mstar\ diagram, above which galaxies are defined as compact SF galaxies. However, the threshold of \sgm\ to separate cSFGs and eSFGs is dependent on stellar mass. We then define a parameter, the deviation of \sgm\ from the demarcation line of cSFGs and eSFGs in logarithmic space ($\Delta$\sgm), to quantify the compactness of the galactic center. The $\Delta$\sgm\ does not depend on stellar mass: cSFGs with $\Delta$\sgm$>0$ and eSFGs with $\Delta$\sgm$<0$. 
Similarly, we use the deviation of SFR from the normal SFMS, $\Delta$SFR, to quantify the star formation status of galaxies, which is independent of stellar mass. 

The left panel of Figure \ref{fig:HI_diagram} presents the  $\Delta$SFR-$\Delta$\sgm\ diagram for all the sample galaxies with the color-coding of \HI\ detection rate. Galaxies on the $\Delta$SFR-$\Delta$\sgm\ diagram are mainly distributed in two branches, denoted by the gray contours. Galaxies in the upper branch are SF galaxies, whereas galaxies in the lower branch are quenched ones. Almost all the quenched galaxies have high inner surface stellar mass densities, indicated by the lack of galaxies at the bottom left corner in Figure \ref{fig:HI_diagram}. This agrees with the previous finding that compact stellar core is the requisite for quenching a galaxy \citep[e.g.][]{Fang-13, Bluck-14, Barro-17}.  
More importantly, from eSFGs to cSFGs, and from cSFGs to quenched population, the \HI\ detection rate is significantly decreasing, which is consistent with the compaction and quenching scenario that the gas reservoirs for sustaining star formation become less and less along the quenching track.  Moreover, we have checked the $\Delta$SFR-$\Delta$\sgm\ diagram with color-coding of the \HI\ mass and find that the result strengthens the above statement. 

In the right panel of Figure \ref{fig:HI_diagram}, we present the sketch for galaxy evolution on the  $\Delta$SFR-$\Delta\Sigma_{\rm 1kpc}$ diagram under the scenario of compaction and quenching, which roughly demonstrates what happened to galaxies during the processes of compaction and quenching. From eSFGs to cSFGs, galaxies appear to significantly consume or loss their cold gas but sustain or even slightly enhance their star formation activities along with metal enhancement (for less massive galaxies) and central stellar mass assembly (see Figure \ref{fig:sfr_zgas} and Figure \ref{fig:sersic}).  From cSFGs to quenched population, galaxies appear to exhaust almost all their cold gas and shut down their star formation along with the build-up their dense stellar cores or bulges (see Figure \ref{fig:HI_diagram} and \ref{fig:sersic}). 

Thus, we propose a simple picture under the scenario of compaction and quenching to string the results together, where eSFGs first contract via a series of possible physical processes.
The dominated processes for compaction are likely to be different at different redshifts. In high redshift universe ($z>1$), major mergers and the accretion of counter-rotating tidal streams occur more frequently than local universe \citep[e.g.][]{Conselice-06, Bertone-Conselice-09, Zolotov-15, Tacchella-16a}. These processes are usually associated with violent disk instabilities, subsequently leading to the collapse of gaseous and/or stellar disk and the enhancement of star formation. However, the physical processes of compaction in local universe are much gentler and have longer timescale than those in high redshift.  Minor mergers, interactions with close companions and/or the bar-driven gas inflow in secular evolution are thought to play a role in enhancing star formation and contributing to the build-up of stellar cores or bulges for low-redshift galaxies \citep[e.g.][]{Moore-Lake-Katz-98, DiMatteo-07, Bournaud-Jog-Combes-07, Wang-12, Lin-17}.
This intense star formation could only sustain for 1 Gyr or less (see the right panel of Figure \ref{fig:sfr_scatter}), but it is efficient to consume the cold gas and assemble the  stellar mass by in-situ star formation in the inner regions of galaxies along with the metal enhancement. 
The major role of compaction is to transform an extended SF galaxy with sufficient \HI\ reservoir to a compact SF but \HI-deficient galaxy (see Figure \ref{fig:map_HI} and Figure \ref{fig:sersic}). 
Then this galaxy can be quenched by either the feedback of its central massive black hole, or stabilizing the gaseous disk against collapse from forming stars because of the existence of a dominant core (or bulge), although the morphological quenching picture appears to be inconsistent with the fact that quenching inevitably invokes the loss of cold gas (see Figure \ref{fig:HI_diagram}). 
Alternatively, this galaxy can be quenched by the environmental effects, such as tidal stripping or strangulation if it is a less massive galaxy of big cluster. We note that the cSFGs and eSFGs appear to reside in similar environment, suggesting that the compaction process is independent of environmental effects. 

We also remind the readers that there is no necessity for invoking the evolution from eSFGs and cSFGs to explain our result. It is also possible that the eSFGs and cSFGs are formed to be different, due to different formation and gas accretion histories. 
By investigating 100 simulated galaxies of various morphologies with Milky-Way-like halo mass drawn from the Galaxies-Intergalactic Medium Interaction Calculation (GIMIC), \cite{Sales-12} found that the coherent alignment of the angular momentum of accreted gas over galaxy formation is the key to galaxy morphology:  spheroids more likely form when the spin of newly accreted gas is misaligned with that of the extant galaxy, whereas disk galaxies form usually with the similar alignment of angular momentum for the accreted gas and the earlier accreted material.  Furthermore, under this scenario, the accreted gas flows more easily into the galactic center for spheroids than disk galaxies, associated with the enhancement of central star formation activities \citep{Davis-11, Chen-16, Jin-16} and the possible feeding of central massive black holes  (see Figure \ref{fig:agn_activity}).  This naturally leads to the more \HI-deficient of cSFGs than eSFGs due to the rapid cold gas consumption in cSFGs, and the structural dependence of metallicity can be caused either by the metal-poor gas accretion of eSFGs or the metal enhancement of rapid star formation activities in the center of cSFGs \citep{Yabe-12, Wang-17, SanchezAlmeida-DallaVecchia-18}. 
However, it is also possible that eSFGs accrete cold gas and/or dwarf galaxies with mis-aligned angular momentum, 
and subsequently turn to cSFGs. 

\section{Summary and Conclusion}
\label{sec:summary}

In this work, we select a volume-limited galaxy sample to have $0.02<z<0.05$, \lgmstar$>$9.5 and minor-to-major axis ratio greater than 0.5 from NSA galaxies located in the ALFALFA footprint. We then match the selected galaxies with MPA-JHU catalog, which results in 15,933 galaxies as our final sample. Further, we classify the galaxy sample into SF and quenched population according to the SFR-\mstar\ relation, and classify the SF galaxies into cSFGs and eSFGs based on the \sgm-\mstar\ diagram. 
We investigate and compare the properties of cSFGs and eSFGs, including the SFR, gas-phase metallicity, \HI\ content and environment. This is helpful to examine the possible evolution from eSFGs to cSFGs with naturally assuming that galaxies evolve along the track of cold gas consumption and metal enhancement.
The main results are listed below. 

\begin{itemize}

\item  The cSFGs  on average  exhibit  similar  or slightly higher SFR with respect to eSFGs at fixed stellar mass, while the compact SF galaxies dominate the galaxy population with most intense star formation activities (0.5 dex above the normal SFMS).

\item  The cSFGs appear to have higher gas-phase metallicity at low stellar mass end than eSFGs, while the metallicity difference between the two populations vanishes at stellar mass higher than $\sim10^{10.5}$\msolar. 

\item The \HI\ detection rate of cSFGs is significantly lower than that of eSFGs at given stellar mass, implying that cSFGs are more \HI-deficient than eSFGs. Indeed, this result is confirmed by adopting  the estimated \HI\ mass for \HI\ undetected galaxies in ALFALFA. 

\item The cSFGs and eSFGs of similar stellar mass appear to reside in similar environment, quantified by the host halo mass and local overdensity.  

\item  The cSFGs are more frequently to host an AGN than eSFGs at given stellar mass, suggesting that the feeding of central massive black hole is more efficient in cSFGs than in eSFGs. 

\end{itemize}

We examine the \HI\ depletion time of cSFGs and eSFGs, and find that cSFGs are more close to be quenched than eSFGs with assuming the current SFR and neglecting the replenishment of cold gas. All these findings support the compaction and quenching scenario, when assuming that galaxies evolve following the track of cold gas consumption and metal enhancement. 
Under this scenario, the compaction is essential to the build-up of stellar cores (or bulges) along with strong cold gas consumption and metal enhancement. In addition, the environmental effects appear to do not play a key role in the compaction process.  
We also note that there is no necessity of the evolution from eSFGs and cSFGs, since the two populations can be formed to be different due to different formation and gas accretion histories. 
 
\acknowledgments

We thank the anonymous referee for constructive suggestions.
This work is supported by the National Basic Research Program of China (973 Program)(2015CB857004), 
the National Natural Science Foundation of China (NSFC, Nos. 11320101002, 11522324, 11421303, 11703092 
and 11433005) and the Fundamental Research Funds for the Central Universities.
EW acknowledges the support from the Youth Innovation Fund by University of 
Science and Technology of China (No. WK2030220019) and China Postdoctoral Science Foundation funded project (No. BH2030000040). ZP acknowledges the support from the Natural Science Foundation of Jiangsu Province (No.BK20161097). 


\bibliography{rewritebib.bib}

\appendix

\subsection{{\rm A}. Using $\Sigma_{\rm e}$ to define samples}
\label{subsec:appendix_a}

\begin{figure*}
  \begin{center}
   \epsfig{figure=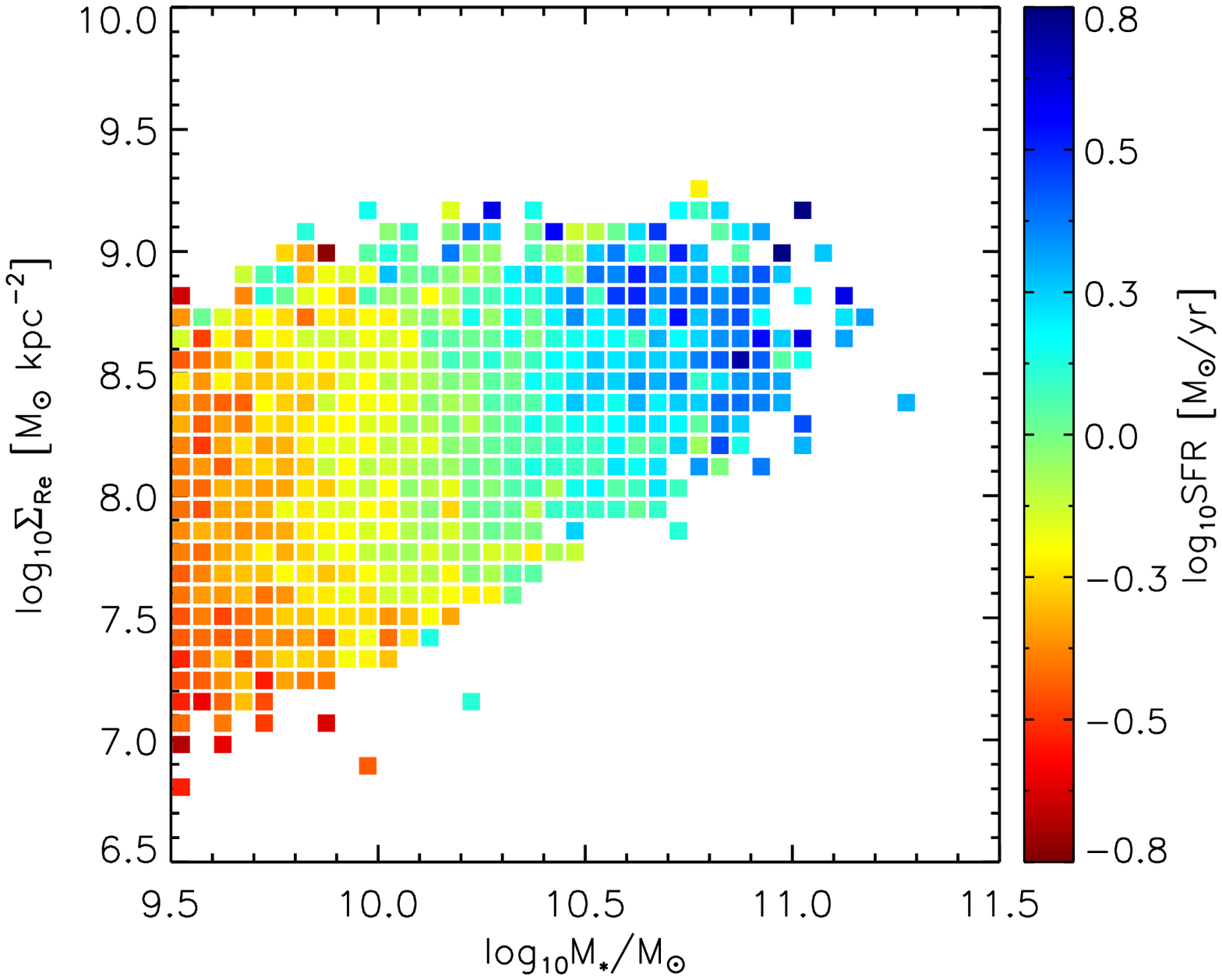,clip=true,width=0.33\textwidth}
   \epsfig{figure=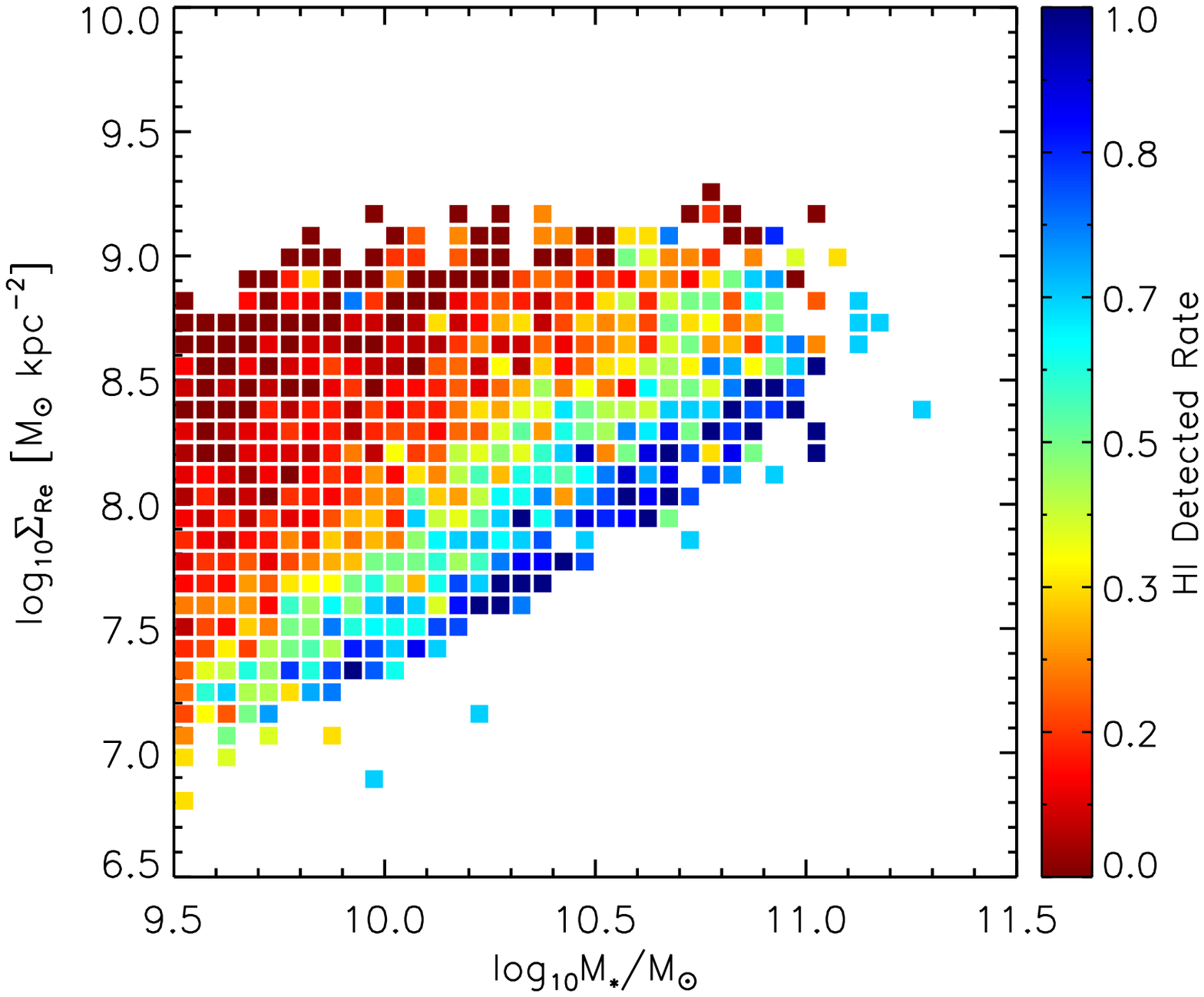,clip=true,width=0.33\textwidth}
   \epsfig{figure=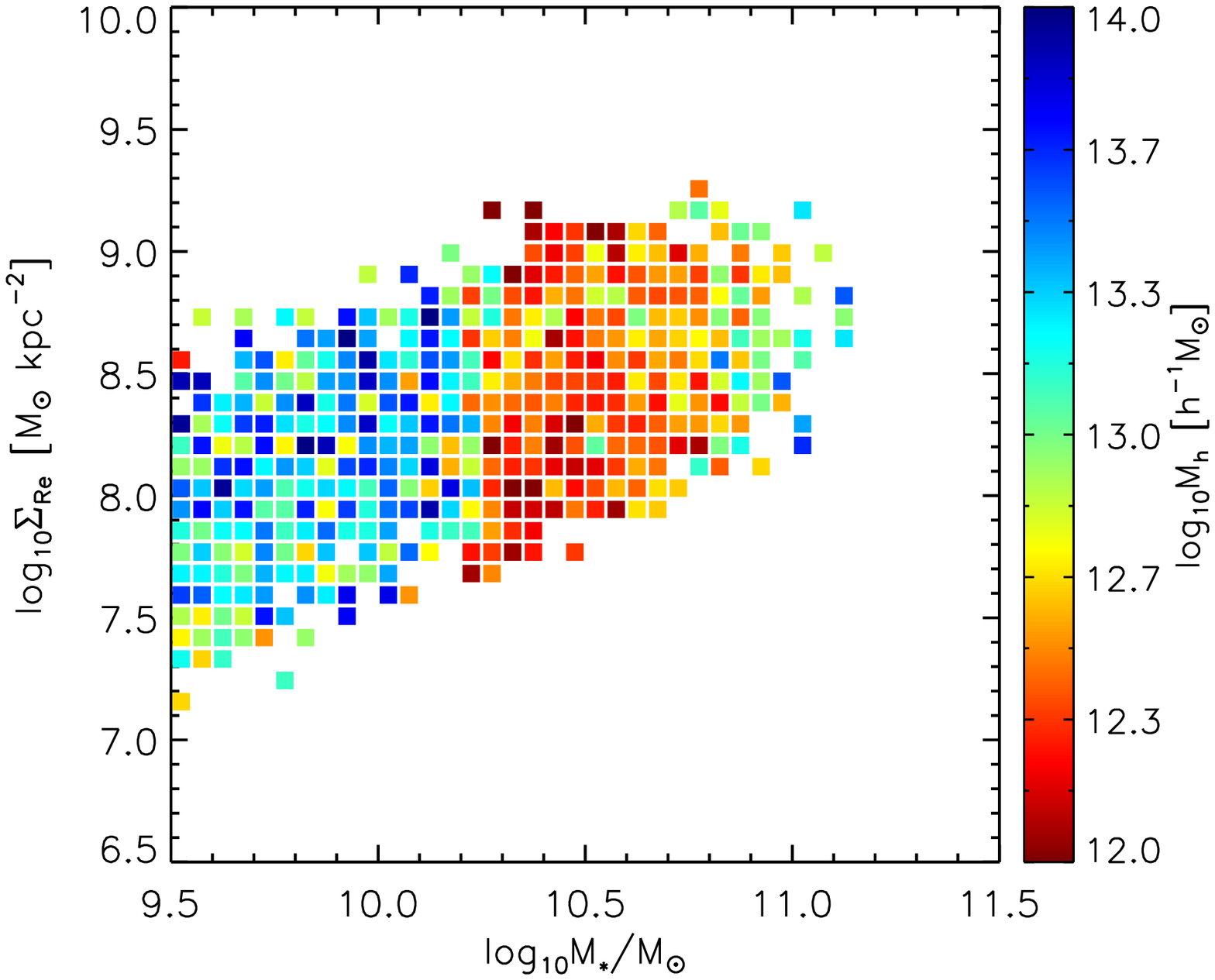,clip=true,width=0.33\textwidth}

   \epsfig{figure=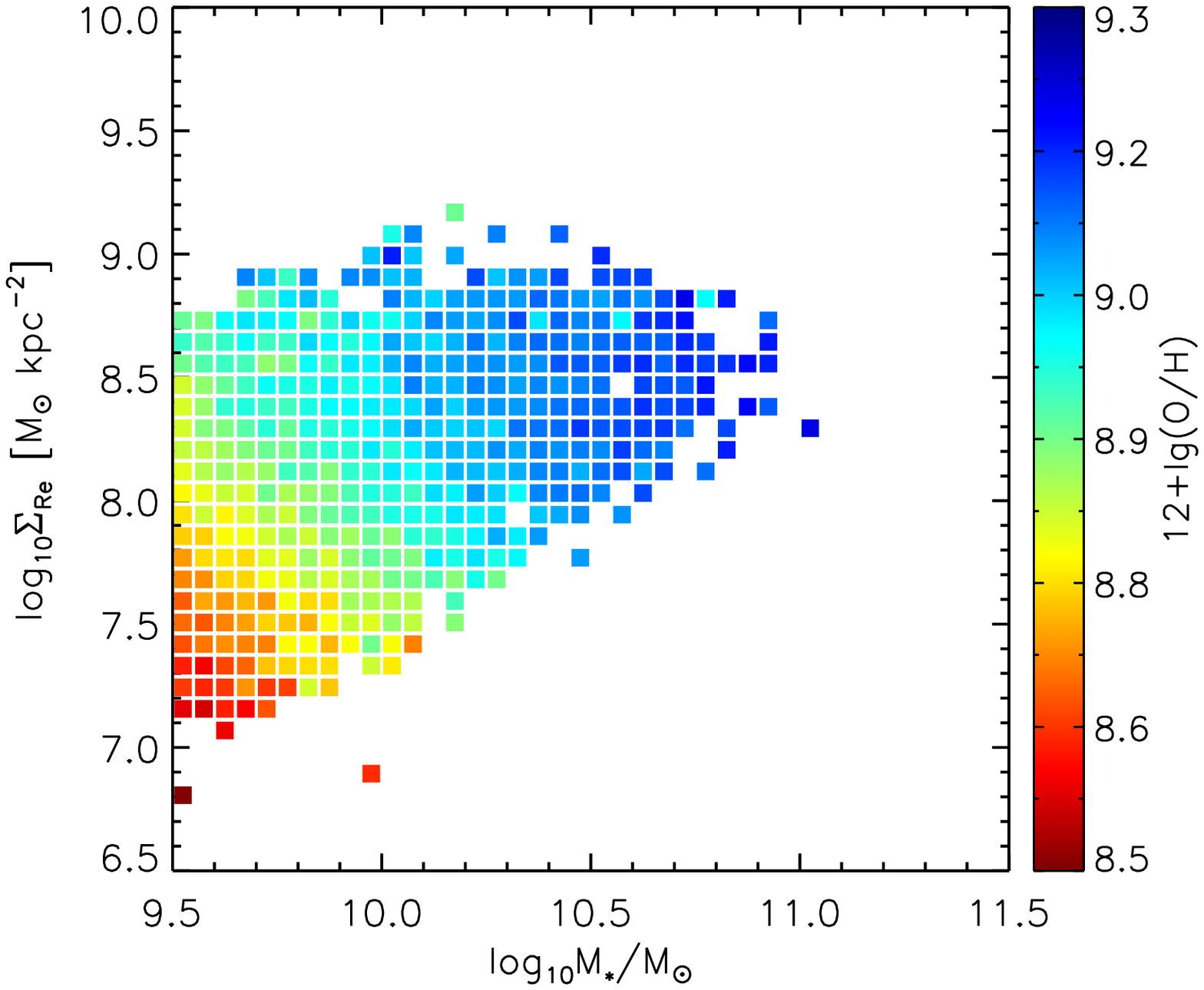,clip=true,width=0.33\textwidth}
   \epsfig{figure=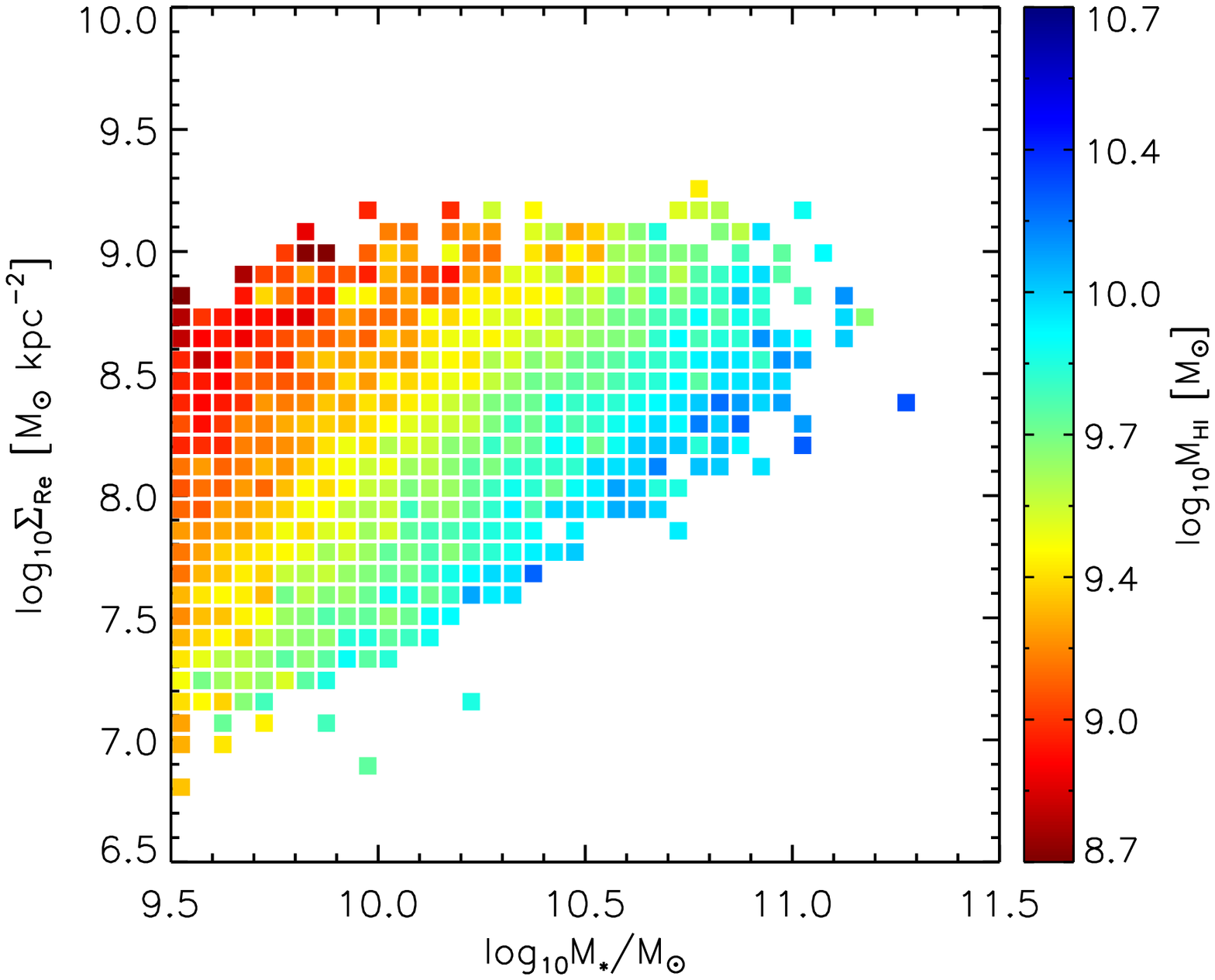,clip=true,width=0.33\textwidth}
   \epsfig{figure=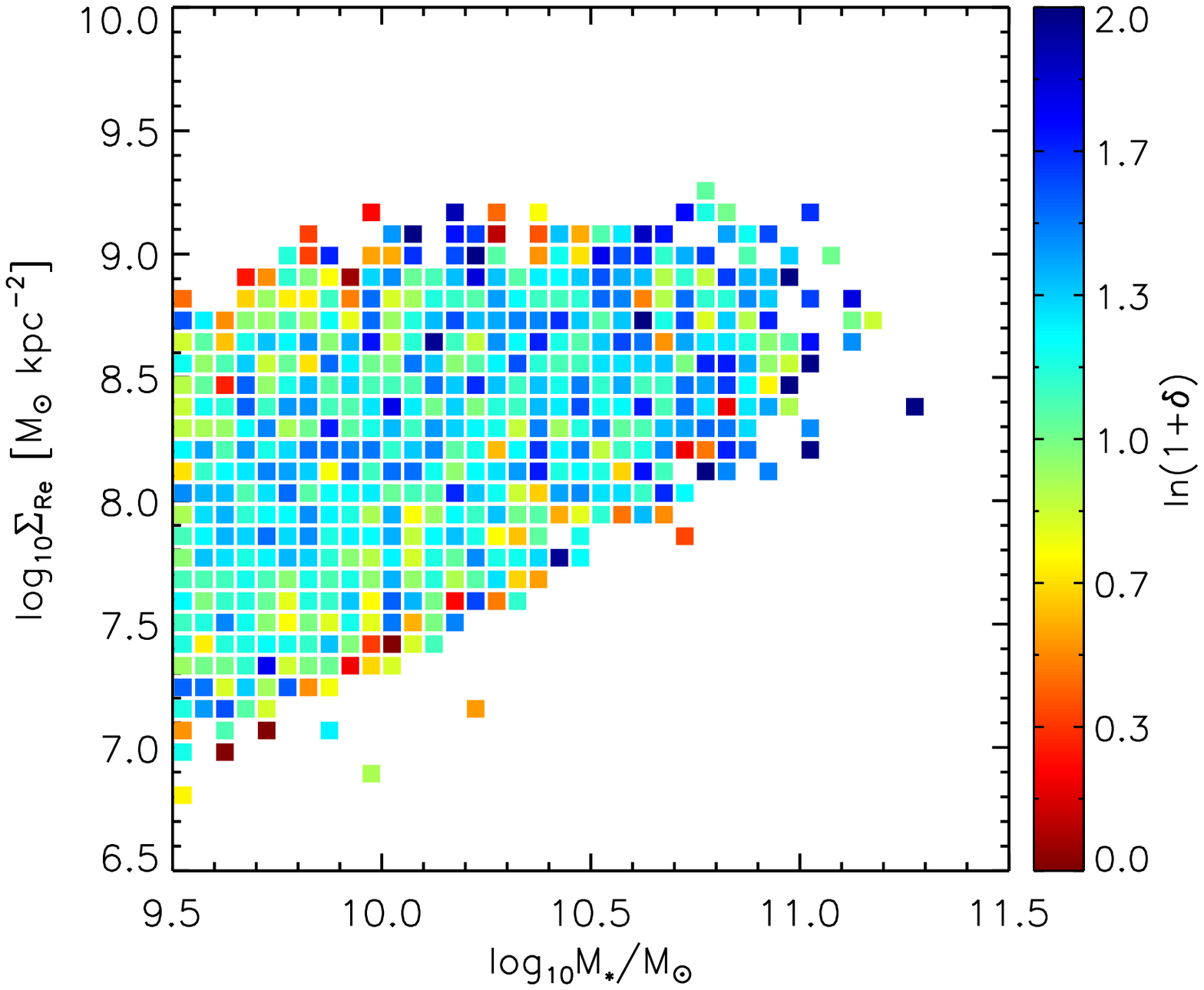,clip=true,width=0.33\textwidth}
  \end{center}
\caption{The $\Sigma_{\rm Re}$-\mstar\ diagram for SF galaxies color-coded with SFR, gas-phase metallicity, the \HI\ detection rate, \HI\ mass, host halo mass and local overdensity from up to bottom and from left to right, respectively.   }
 \label{fig:map_re}
\end{figure*}

In the main text, we use the surface mass density within a radius of 1 kpc to quantify the compactness of galaxies, and to classify the cSFGs and eSFGs. However, one may worry that the aperture of 1 kpc corresponds to different relative size of different galaxies. Here we present another parameter, the stellar mass surface density within the effective radius, $\Sigma_{\rm Re}$, to reproduce the main result of this work.  In Figure \ref{fig:map_re}, we present the $\Sigma_{\rm Re}$-\mstar\ relation with the color-coding of SFR, gas-phase metallicity, \HI\ detection rate, \HI\ mass, the host halo mass and the local overdensity. As shown, galaxies in the $\Sigma_{\rm Re}$-\mstar\ diagram show broader distribution than in \sgm-\mstar\ diagram, which is consistent with previous studies \citep[e.g.][]{Fang-13, Barro-17}. We find that the SFR and environmental properties show weak or no dependence on $\Sigma_{\rm Re}$, while the \HI\ content and gas-phase metallicity show significant dependence on $\Sigma_{\rm Re}$. This is in good agreement with our result, indicating that the result is not sensitive to the definition of cSFGs and eSFGs.  

\subsection{{\rm B}. \sgm-\mstar\ diagram with color-coding of \sersic\ index}
\label{subsec:appendix_b}
\begin{figure*}
  \begin{center}
   \epsfig{figure=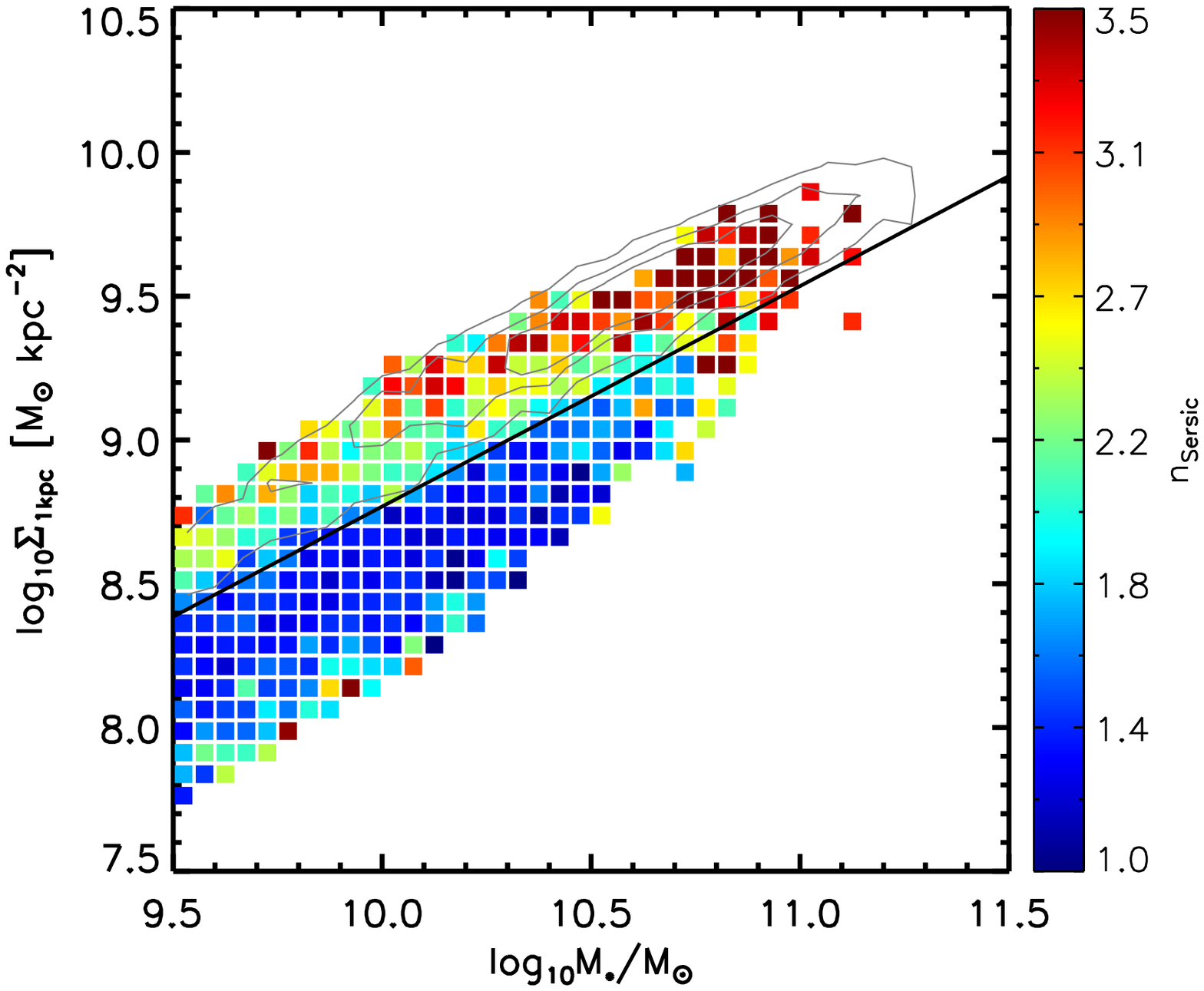,clip=true,width=0.405\textwidth}
   \epsfig{figure=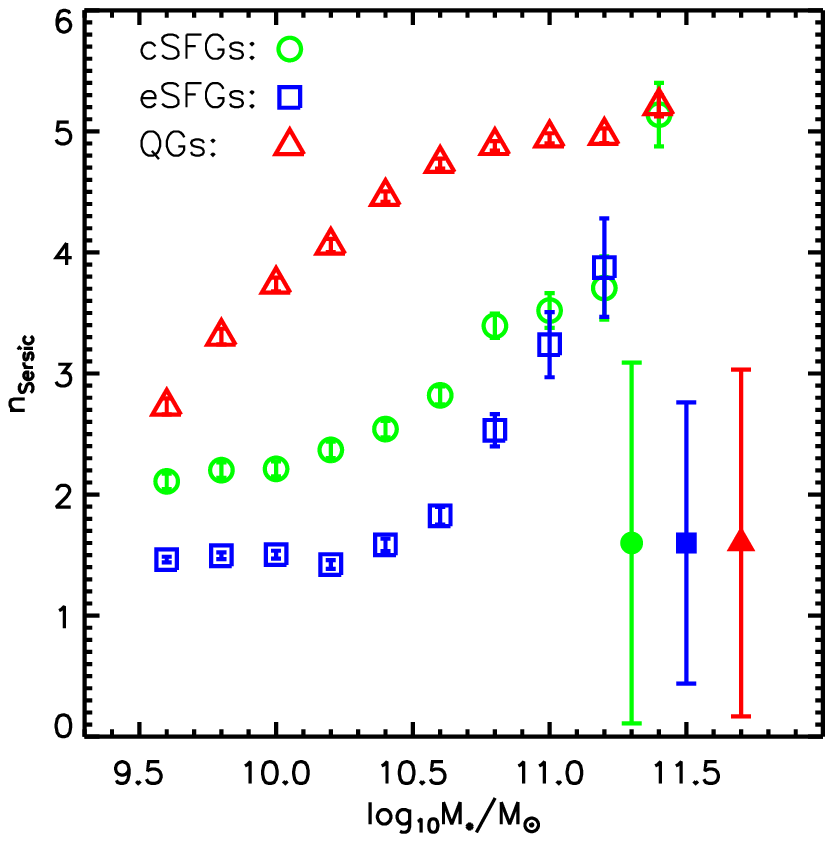,clip=true,width=0.378\textwidth}
  \end{center}
\caption{Left panel: the \sgm-\mstar\ diagram for SF galaxies color-coded with \sersic\ index. Right panel: the \sersic\ index as a function of stellar mass for eSFGs (blue squares), cSFGs (green circles) and quenched galaxies (red triangles). }
\label{fig:sersic}
\end{figure*}

Previous results have shown that the \sersic\ index is increasing from the eSFGs, through cSFGs to quenched galaxies \citep{Barro-17}. Here we present the \sgm-\mstar\ for SF galaxies with the color-coding of \sersic\ index in the left panel of Figure \ref{fig:sersic}.  The \sersic\ indices are taken from NSA catalog \citep{Blanton-11}. 
As shown, the \sersic\ index strongly depends on the \sgm, with the higher \sgm\ the higher \sersic\ index. In addition, the demarcation line of cSFGs and eSFGs appear to perfectly distinguish the \sersic\ index of sample galaxies, in the sense that, galaxies below and above the demarcation line show clear different \sersic\ index. This confirms that our classifications of cSFGs and eSFGs are reasonable.  
The right panel of Figure \ref{fig:sersic} shows the \sersic\ index as a function of stellar mass for cSFGs, eSFGs and quenched galaxies. The \sersic\ indices of cSFGs are between those of eSFGs and quenched galaxies almost over the whole stellar mass range. This is consistent with the compaction and quenching scenario that the structural properties evolve from eSFGs, through cSFGs to quenched galaxies along with the morphological transformation from disk-like to spheroid-like.

\label{lastpage}
\end{document}